\documentclass[10pt, a4paper, onecolumn, conference, compsocconf]{IEEEtran}

\usepackage{graphicx}
\usepackage{url}
\usepackage{times}
\usepackage{color}
\usepackage{amssymb,amsmath}
\usepackage{threeparttable}
\usepackage[ruled,vlined]{algorithm2e}
\usepackage{threeparttable}
\newsavebox{\tablebox}

\newcommand{\true}{\texttt{True}}
\newcommand{\false}{\texttt{False}}

\newcommand\bool{\mathbb{B}}
\newcommand\boolform{{\cal B}}
\newcommand\eval{{\cal E}}
\newcommand{\initloc}[1]{l^0_{#1}}
\newcommand{\initeval}[1]{e^0_{#1}}

\newcommand{\system}{\mathcal{S}}
\newcommand{\Vis}{\textsf{Vis}}

\newcommand{\reach}{\mathcal{R}_{\system}\xspace}
\newcommand{\conf}{\mathcal{C}_{\system}\xspace}

\usepackage{times}

\newenvironment{list1}{\begin{list}{$\bullet$}
{\topsep 0 pt \parsep 0 pt \partopsep 0 pt \itemsep 0
pt}}{\end{list}}

\usepackage{xspace}
\newcommand{\Attr}{\mathsf{Attr}\xspace}

\newcommand{\sys}{\mathcal{S}}

\newtheorem{defi}{Definition}
\newtheorem{theo}{Theorem}
\newtheorem{prop}{Proposition}
\newtheorem{assump}{Assumption}

 \newcommand{\comment}[1]{}
\begin{document}
%
\title{Distributed Priority Synthesis and its Applications}


\author{
\IEEEauthorblockN{Chih-Hong Cheng\authorrefmark{1}\authorrefmark{2},
                    Saddek Bensalem\authorrefmark{3},
                    Rongjie Yan\authorrefmark{4},}
\IEEEauthorblockN{Harald Ruess\authorrefmark{2},
                    Christian Buckl\authorrefmark{2},
                    Alois Knoll\authorrefmark{1}}
                    \\
\IEEEauthorblockA{\authorrefmark{1}Department of Informatics, Technische Universit\"{a}t M\"{u}nchen, Munich, Germany}
\IEEEauthorblockA{\authorrefmark{2}fortiss GmbH, Munich, Germany}
\IEEEauthorblockA{\authorrefmark{3}Verimag Laboratory, Grenoble, France}
\IEEEauthorblockA{\authorrefmark{4}State Key Laboratory of Computer Science, ISCAS, Beijing, China}
\texttt{http://www.fortiss.org/formal-methods}
}


\maketitle

\begin{abstract}
Given a set of  interacting components with non-deterministic variable update and given  safety requirements,
the goal  of {\em priority synthesis} is to restrict, by means of priorities,  the set of possible interactions in
 such a way as to guarantee the  given safety conditions for all possible runs.
 In {\em distributed priority synthesis} we are interested in obtaining local  sets of priorities, which
are deployed in terms of local component controllers sharing  intended next moves between  components in local neighborhoods only.
These possible communication paths between local controllers are specified  by means of a  {\em communication architecture}\@.
 We formally define the problem of distributed priority synthesis in terms of a multi-player safety game between  players for (angelically) selecting the next transition of the components and an environment for (demonically) updating uncontrollable variables; this problem is NP-complete.
We propose several optimizations including a  solution-space exploration  based on  a diagnosis method using   a nested extension
 of the usual attractor computation in games together with a reduction to corresponding  SAT problems.
When diagnosis fails, the method proposes potential candidates to guide the exploration.
These optimized algorithms for solving  distributed priority synthesis problems have been integrated into our  VissBIP framework.
An experimental validation of this implementation  is performed  using  a range  of  case studies including scheduling in multicore processors
 and modular robotics.
\end{abstract}

\section{Introduction\label{sec.dps.introduction}}

Given a set of  interacting components with non-deterministic variable update and given a  safety requirement on the overall system,
the goal  of {\em priority synthesis} is to restrict, by means of priorities on interactions,  the set of possible interactions in
 such a way as to guarantee the  given safety conditions.
Since many well-known scheduling strategies
can be encoded by means of priorities on interactions~\cite{goessler2003priority}, priority synthesis is closely related to solving scheduling problems.

\begin{figure}[t]
    \centering
     \includegraphics[width=0.9\columnwidth]{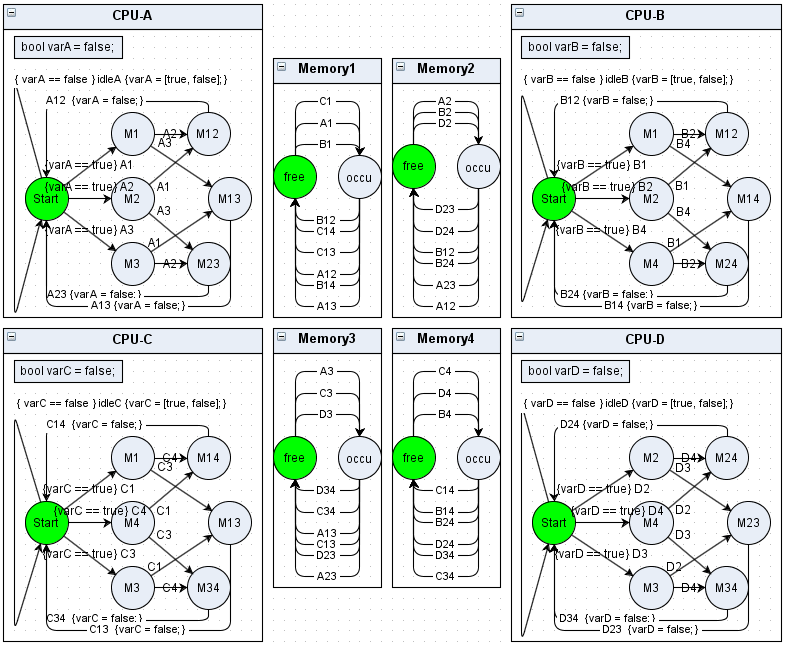}
      \caption{Multicore scheduling in VissBIP~\cite{cheng:vissbip:2011}.}
     \label{fig:VissBIP.Construction}
\end{figure}

Consider, for example,  the multiprocessor scheduling  scenario depicted  in Figure~\ref{fig:VissBIP.Construction} as motivated by a 3D image processing application.
Each of the four processors needs to allocate two out of four  memory banks for processing; in this model processor {\tt A} (in state {\tt Start}) may
 allocate memory bank {\tt 2} (in state {\tt free})  by synchronizing on the transition with label {\tt A2}, given that CPU  {\tt A} is ready to process -  that is {\tt  varA}, which is non-deterministically toggled  by the  environment through {\tt idleA} transitions, holds.
 Processor {\tt A}  may only allocate its "nearest" memory banks  {\tt 1}, {\tt 2} and {\tt 3}\@.
Without any further restrictions on the control this multiprocessor system  may deadlock.

Such control restrictions are expressed in terms of priorities between possible  interactions.
For instance, a priority {\tt B1 < A1} effectively disables interaction {\tt B1} whenever {\tt A1} is enabled.
 A solution for the priority synthesis problem, based on game-theoretic notions and a translation to a
corresponding satisfiability problem, has been described previously~\cite{cheng:vissbip:2011,cheng:algo.priority.syn:2011}\@.
This solution yields centralized controllers, whereas here we are interested in obtaining decentralized controls for each of the components. Coordination between these local controllers is restricted to communicating intended next moves along predefined communication paths.

The possible communication paths among components are defined in terms of a {\em communication architecture} which consists of ordered pairs of components\@. For example, executing interaction {\tt A2} requires bidirectional communications along {\tt (A,M2)} and  {\tt (M2,A)}.
A master-slave communication architecture for broadcasting the next transition of processor {\tt A}  to all other processors  includes pairs {\tt (A,B)},
{\tt (A,C)},  and {\tt (A,D)}.
In this architecture (Table~\ref{table:VissBIP.result}: index~1), the local controller for each of the recipient CPUs uses the communicated next transition of CPU {\tt A}, say {\tt A1}, and disables every enabled
local transition with a lower priority than {\tt A1}\@.
Alternative architectures in Figure~\ref{table:VissBIP.result} for the multiprocessor scenario include a two-master protocol where processors  {\tt A} and {\tt D} notify processors {\tt B} and {\tt C}, and a symmetric architecture where each of the processors notifies its "nearest" neighbor. Notice that communication architectures are not necessarily transitive.

Altogether, the result of  {\em distributed priority synthesis} are certain sets of local priorities for each component which are compatible with a given communication architecture.
More precisely, if component $C$ may notify component $D$ in a given communication architecture, then local priorities for the controller of component $D$ are of the form $s < t$,
where $s$ is a possible transition of $D$ and $t$ a possible transition of $C$\@.
Possible solutions for three different communication architectures for the multiprocessor scenario are listed in Table~\ref{table:VissBIP.result}\@.
Notice that the solution for the symmetric architectures (index~3) uses a slight refinement in that components do not only publish the intended next transition but also the source state of this transition; for example, the notation {\tt A1.M2} expresses that processor {\tt A} is at location {\tt M2} and intends to trigger transition {\tt A1}\@.
Obviously, this refined notion of priorities can always be expressed in a transformed model with new transitions, say {\tt A1M2},  {\tt A1M3}, {\tt A1Start},
for encoding the source states {\tt M2},  {\tt M3}, {\tt Start} of {\tt A1}\@.

Given a solution to the distributed priority synthesis problem, a local controller for each component may work in each cycle by, first,  sending its intended next move and receiving next moves from other components according to the given communication architecture, and, second, disabling any enabled local transitions with
 a lower priority among the received intended next moves; algorithms for priority deployment~\cite{Bonakdarpour2011distribute,bensalem2010methods} may be reused.

The rest of the paper is structured as follows.
Section~\ref{sec.dps.formulation} contains background information on a simplified variant of the Behavior-Interaction-Priority (BIP)
modeling framework~\cite{basu2006modeling}\@. The  corresponding  priority synthesis problem corresponds to  synthesizing a state-less winning strategy
in a two-player safety game, where  the control player (angelically) selects the next transition of the components and the  environment player  (demonically)
updates uncontrollable variables.
In Section \ref{sec.algo.prioritysyn.distributed.execution} we introduce the notion of deployable communication architectures and formally state the
 distributed priority synthesis problem.
 Whereas the general distributed controller synthesis problem is undecidable~\cite{PnueliFOCS90} we show that distributed priority synthesis is NP-complete\@.
 Section~\ref{sec.dps.algorithm} contains a solution to the distributed synthesis problem, which is guaranteed to be deployable on a given communication architecture.
This algorithm is a generalization of the solution to the priority synthesis problem in~\cite{cheng:vissbip:2011,cheng:algo.priority.syn:2011}\@. It is a complete algorithm and
integrates essential optimizations based on symbolic game encodings including visibility constraints, followed by a nested attractor computation, and lastly, solving a corresponding (Boolean) satisfiability
 problem by extracting fix candidates while considering architectural constraints.
 Section~\ref{sec.dps.algo.extension}  describes some details and optimization of our implementation,
which is validated in Section~\ref{sec.dps.evaluation} against  a set of selected case studies including scheduling in multicore processors and modular robotics.
 Section~\ref{sec.dps.related.work} contains related work and we  conclude  in  Section~\ref{sec.dps.conclusion}.



\begin{table}[t]
\centering
\begin{scriptsize}
\begin{lrbox}{\tablebox}
\begin{threeparttable}
\begin{tabular}{|l|p{2.6cm} |p{2.3cm} |p{2.3cm} |p{2.3cm} |p{2.3cm} |}
\hline
 & Additional Communication  & Controller A & Controller B & Controller C & Controller D \\
\hline
1   & /* A broadcast to B, C, D */ & unrestricted  & $(B1 < A1)$ & $(C1 < A1)$  & $(D2 < A2) $ \\
   &   (CPU-A, CPU-B)    &               & $(B2 < A2)$ & $(C3 < A3) $  & $(D3 < A3)$ \\
   &   (CPU-A, CPU-C)    &               & $(B1 < idleA)$ & $(C1 < idleA)$  & $(D2 < idleA)$ \\
   &   (CPU-A, CPU-D) &               & $(B2 < idleA)$ & $(C3 < idleA)$  & $(D3 < idleA)$ \\
\hline
2   & /* A, D send to B, C */                        & unrestricted  & $(B1 < A1)$   & $(C1 < A1)$ & unrestricted  \\
   &   (CPU-A, CPU-B)    &               & $(B1 < idleA)$ & $(C1 < idleA)$  &  \\
   &   (CPU-A, CPU-C)   &               & $(B2 < A2)$ & $(C3 < A3)$  & \\
   &      (CPU-D, CPU-B)                &               & $(B2 < idleA)$ & $(C3 < idleA)$  & \\
      &     (CPU-D, CPU-C)                                           &               & $(B4 < D4)$ & $(C4 < D4)$  & \\
      &                                              &               & $(B4 < idleD)$ & $(C4 < idleD)$  & \\
      &                                              &               & $(idleB < A1)$ & $(idleC < A1)$  & \\
      &                                              &               & $(idleB < A2)$ & $(idleC < A2)$  & \\
      &                                              &               & $(idleB < A3)$ & $(idleC < A3)$  & \\
      &                                              &               & $(idleB < D4)$ & $(idleC < D4)$  & \\
      &                                              &               & $(idleB < idleA)$ & $(idleC < idleA)$  & \\
\hline
3   & /* local communication */ &  $(A1.St < B1.M2)$             &  $(B1.St < A1.M2)$ & $(C1.St < A1.M2)$ & $(D2 < B2.M1)$  \\
      &   (CPU-A, CPU-B)     &  $(A1.St < B1.M4)$             &  $(B1.St < A1.M3)$ & $(C1.St < A1.M3)$ & $(D2 < B2.M4)$ \\
      &   (CPU-A, CPU-C)   &  $(A2.St < B2.M1)$             &  $(B2.St < A2.M1)$ & $(C3.St < A3.M1)$ & $(D3.St< C3.M1)$ \\
      &   (CPU-B, CPU-A)    &  $(A2.St < B2.M4)$             &  $(B2.St < A2.M3)$ & $(C3.St < A3.M2)$ & $(D3.St < C3.M4)$ \\
      &   (CPU-B, CPU-D)    &  $(A3.St < C3.M1)$             &  $(B4.St < D4.M2)$ & $(C4.St < D4.M2)$ & $(D4.St < B4.M1)$ \\
      &    (CPU-C, CPU-A)      &  $(A3.St < C3.M4)$             &  $(B4.St < D4.M3)$ & $(C4.St < D4.M3)$ & $(D4.St < B4.M2)$ \\
      &   (CPU-C, CPU-D) (CPU-D, CPU-C) (CPU-D, CPU-B)      &               &  &  &  \\
\hline
\end{tabular}
\end{threeparttable}
\end{lrbox}
\scalebox{0.96}{\usebox{\tablebox}}
\end{scriptsize}
      \caption{Communication structures and corresponding distributed controllers for multiprocessor scenario in Figure~\ref{fig:VissBIP.Construction}.
                     Notice that St abbreviates Start. 
                    }
     \label{table:VissBIP.result}
\end{table}

\section{Background\label{sec.dps.formulation}}

Our notion of {\em interacting components} is heavily influenced by the Behavior-Interaction-Priority (BIP) framework~\cite{basu2006modeling} which
consists of a set of automata (extended with data) that synchronize on joint labels; it is designed to model systems with combinations of synchronous and
asynchronous composition.
For simplicity, however, we omit many syntactic features of BIP such as hierarchies of interactions and we restrict ourselves to Boolean data types only. Furthermore, uncontrollability is restricted to non-deterministic update of variables, and data transfer among joint  interaction among components is also omitted.

Let $\Sigma$ be  a nonempty alphabet of {\em interactions}\@.
A {\em component} $C_i$  of the form $(L_i, V_i, \Sigma_i, T_i, \initloc{i}, \\\initeval{i})$  is a {\em  transition system} extended with data,
where  $L_i$  is a nonempty, finite set of \emph{control locations},
           $\Sigma_i \subseteq \Sigma$ is  a nonempty subset of interaction labels used in $C_i$,  and
            $V_i$  is a finite set of \emph{(local) variables} of Boolean domain  $\bool = \{\true, \false\}$\@.
           The set  $\eval(V_i)$ consists  of all evaluations $e: V_i\to\bool$\ over the variables $V_i$\@, and $\boolform(V_i)$ denotes the set of
          propositional formulas over variables in $V_i$; variable evaluations are extended to propositional formulas in the obvious way\@.
 $T_i$ is the set of \emph{transitions} of the form $(l,g,\sigma,f,l')$, where
                  $l, l'\in L_i$ respectively are the source and target locations,
                  the guard  $g \in \boolform(V_i)$  is a  Boolean formula over the variables $V_i$\@,
                  $\sigma \in \Sigma_i$ is an interaction label (specifying the event triggering the transition), and
                  $f: V_i \rightarrow (2^\bool \setminus \emptyset)$ 
                  is the {\em update relation} mapping every variable to a set of allowed Boolean values.
 Finally, $\initloc{i} \in L_i$ is the \emph{initial location} and $\initeval{i} \in \eval(V_i)$ is the initial evaluation of the variables.

\newcommand{\inter}{\bar{\sigma}}

A system $\mathcal{S}$ of {\em interacting components} is of the form  $(C = \bigcup_{i=1}^m C_i  , \Sigma, \mathcal{P})$, where $m \geq 1$, all the $C_i$'s are components,
the set of {\em priorities} $\mathcal{P} \subseteq 2^{\Sigma\times\Sigma}$ is irreflexive and transitive~\cite{goessler2003priority}.
 The notation  $\sigma_1 \prec \sigma_2$ is usually used instead of $(\sigma_1, \sigma_2) \in \mathcal{P}$, and we say that $\sigma_2$ has higher priority than $\sigma_1$\@.
A \emph{configuration (or state)} $c$ of a system $\mathcal{S}$ is of the form $(l_1, e_1, \ldots, l_m, e_m)$ with $l_i \in L_i$ and $e_i \in \eval(V_i)$ for all $i \in \{1,\ldots,m\}$\@.
The \emph{initial configuration} $c_0$ of   $\mathcal{S}$ is of the form  $(\initloc{1}, \initeval{1}, \ldots, \initloc{m}, \initeval{m})$\@.
An interaction $\sigma \in \Sigma$ is \emph{(globally) enabled} in a configuration $c$  if,
first, joint participation holds for $\sigma$, that is,  for all $\sigma \in \Sigma_i$ with $i \in \{1,\ldots, m\}$, there exists a transition  $(l_i,g_i,\sigma,f_i,l_i') \in T_i$ with  $e_i(g_i) = \true$\@,
and, second, there is no other interaction of higher priority for which joint participation holds.
$\Sigma_c$ denotes the set of (globally) enabled interactions in a configuration $c$\@.
For $\sigma \in  \Sigma_c$,
a configuration  $c'$ of the form $(l'_1, e'_1, \ldots, l'_m, e'_m)$ is a $\sigma$-\emph{successor} of $c$, denoted by  $c \xrightarrow[]{\sigma} c'$, if,
for all $i$ in $\{1,\ldots,m\}$,

\begin{list1}
\item  if $\sigma \not\in \Sigma_i$, then $l'_i = l_i$ and $e'_i = e_i$;
\item  if $\sigma \in \Sigma_i$  and (for some) transition
         of the form $(l_i,g_i,\sigma,f_i,l_i') \in T_i$ with $e_i(g_i) = \true$,
$e'_i= e_i[v_i / d_i]$ with $d_i \in f(v_i)$.
\end{list1}

A  \emph{run} is of the form $c_0,\ldots,c_k$ with $c_0$ the initial configuration and   $c_j \xrightarrow[]{\sigma_{j+1}} c_{j+1}$  for all  $j: 0\le j< k$\@. In this case, $c_k$ is reachable, and  $\reach$  denote the set of all reachable configurations from $c_0$\@.
Notice that such a sequence of configurations can be viewed  as an execution of a two-player game played alternatively between the control \textsf{Ctrl} and the environment \textsf{Env}\@.
In every position, player  \textsf{Ctrl} selects one of the enabled interactions and \textsf{Env} non-deterministically chooses new values for the variables before moving to the next position.
The game is won by \textsf{Env} if \textsf{Ctrl} is unable to select an enabled interaction, i.e., the system is deadlocked, or if  $\textsf{Env}$ is able to drive the run into a
bad configuration from some given set  $\mathcal{C}_{risk} \subseteq \conf$\@.  More formally,  the system is \emph{deadlocked} in configuration
$c$ if there is no $c' \in \reach$ and no $\sigma \in \Sigma_c$ such that  $c \xrightarrow[]{\sigma} c'$, and the set of deadlocked states is denoted by $\mathcal{C}_{dead}$\@.
A configuration $c$ is \emph{safe} if $c \notin \mathcal{C}_{dead}\cup\mathcal{C}_{risk}$,
and a system is  safe if no reachable configuration is unsafe.
\vspace{2mm}
\begin{defi}[Priority Synthesis]
  Given a system $\mathcal{S} = (C , \Sigma, \mathcal{P})$ together with a set $\mathcal{C}_{risk} \subseteq \mathcal{C}_\sys$ of risk configurations,
 $\mathcal{P}_{+} \subseteq\Sigma\times\Sigma$ is a solution to the \emph{priority synthesis problem} if the extended
  system $(C , \Sigma, \mathcal{P}\cup\mathcal{P}_{+})$ is safe, and the defined relation of $\mathcal{P}\cup\mathcal{P}_{+}$ is also irreflexive and transitive.
\end{defi}
\vspace{2mm}
For the product graph induced by system $\system$, let $Q$ be the set of vertices and $\delta$ be the set of transitions.
In a single player game, where \textsf{Env} is restricted to deterministic updates, finding a solution to the priority synthesis problem is NP-complete in the size of $(|Q|+|\delta|+|\Sigma|)$~\cite{cheng:hardness:2011}\@.

\section{Distributed Execution} \label{sec.algo.prioritysyn.distributed.execution}

We introduce the notion of (deployable) communication architecture for  defining distributed execution for a system  $\mathcal{S}$  of interacting components. 
Intuitively, a communication architecture specifies which components exchange information about their next intended move.

\newcommand{\informs}{\leadsto}
\vspace{2mm}
\begin{defi} 
   A {\emph communication architecture} $Com$ for a system $\mathcal{S}$ of interacting components is  a set of ordered pairs of components of the form  $(C_i, C_j)$ for  $C_i, C_j \in C$\@.
   In this case we say that $C_i$  {\em informs} $C_j$ and we use the notation $C_i \informs C_j$\@.
   Such a  communication architecture $Com$ is  {\em deployable} if the following conditions hold for all $\sigma, \tau \in \Sigma$ and $i,j \in \{1,\ldots, m\}$\@:
  \begin{enumerate}
       \item {\em (Self-transmission)} $\forall i\in \{1,\ldots, m\}$, $C_i\informs C_i \in Com$.
    \item {\em (Group transmission)}  If $\sigma \in \Sigma_i \cap \Sigma_j$ then $C_j \informs C_i,~C_i \informs C_j \in Com$\@.
    \item {\em  (Existing priority transmission)} If $\sigma \prec \tau \in \mathcal{P}$,  $\sigma \in \Sigma_j$, and $\tau\in \Sigma_i$ then $C_i \informs C_j \in Com$\@.
  \end{enumerate}
\end{defi}
\vspace{2mm}

Therefore, components that possibly participate in a joint interaction exchange information about next intended moves (group transmission), and components with a high priority interaction $\tau$ need to inform all components with an interaction of lower priority than $\tau$ (existing priority transmission)\@.
We make the following assumption.

\vspace{2mm}
\begin{assump}[Compatibility Assumption]
It is assumed that a system is deployable on the given communication architecture.
\end{assump}
\vspace{2mm}

Next we define distributed notions of enabled interactions and behaviors, where all the necessary information is communicated along the defined communication architecture.
\vspace{2mm}
\begin{defi}
Given a communication architecture $Com$ for a system $\mathcal{S}$, an interaction $\sigma$ is \emph{visible} by $C_j$ if $C_i \informs C_j$ for all $i$ such that $\sigma\in\Sigma_i$\@.
Then for configuration $c=(l_1, e_1, \ldots, l_m, e_m)$, an interaction $\sigma \in \Sigma$ is \emph{distributively-enabled (at $c$)} if ($i \in \{1,\ldots, m\}$):
\begin{enumerate}
    \item (Joint participation: distributed version)
    for all $i$ with $\sigma \in \Sigma_i$, $\sigma$ is visible by $C_i$, there exists  $(l_i,g_i,\sigma,\_,\_) \in T_i$ with  $e_i(g_i) = \true$\@.
    \item (No higher priorities enabled: distributed version)
     for all  $\tau \in \Sigma$ with  $\sigma \prec \tau$, $\tau$ is visible by $C_i$, and
    there is a $j\in \{1,\ldots, m\}$ such that $\tau \in \Sigma_j$ and either $(l_j,g_j,\tau,\_,\_) \not\in T_j$ or for every $(l_j,g_j,\tau,\_,\_) \in T_j$, $e_j(g_j) = \false$\@.
\end{enumerate}
\end{defi}
\vspace{2mm}
A configuration $c' = (l_1', e_1', \ldots,$ $l_m', e_m')$  is a {\em distributed $\sigma$-successor} of $c$ if $\sigma$ is distributively-enabled and $c'$ is a $\sigma$-successor of $c$.
 {\emph Distributed runs} are runs of system $\mathcal{S}$ under communication architecture $Com$\@.

Any move from a configuration to a successor configuration in the distributed semantics can be understood as a multi-player game with $(|C|+1)$ players between
controllers $\textsf{Ctrl}_i$ for each component and the external environment $\textsf{Env}$\@.
In contrast to the two-player game for the global semantics, $\textsf{Ctrl}_i$ now is only informed on the intended next moves of the components in the
visible region as defined by the communication architecture, and the control players play against the environment player.
First, based on the visibility, the control players agree (cmp. Assumption~\ref{assumption2} below) on an interaction $\sigma \in\Sigma_c$, and, second, the
environment chooses a $\sigma$-enabled transition for each component $C_i$ with $\sigma \in \Sigma_i$\@.
Now the successor state is obtained by local updates to the local configurations for each component and variables are non-deterministically
toggled by the environment\@.

\vspace{2mm}
\begin{prop}\label{prop.distributively.enableness}
Consider a system $\mathcal{S}=(C , \Sigma, \mathcal{P})$ under a deployable communication architecture $Com$. \emph{(a)} If $\sigma \in \Sigma$ is globally enabled at configuration $c$, then $\sigma$ is distributively-enabled at $c$. \emph{(b)} The set of distributively-enabled interactions at configuration $c$ equals $\Sigma_c$. \emph{(c)} If configuration $c$ has no distributively-enabled interaction, it has no globally enabled interaction. 
\end{prop}

\noindent \begin{proof}
(a) An interaction $\sigma \in \Sigma$ is globally enabled in a configuration $c$  if,
first, joint participation holds for $\sigma$, that is,  for all $i \in \{1,\ldots, m\}$ and  $\sigma \in \Sigma_i$  there is a transition  $(l_i,g_i,\sigma,f_i,l_i') \in T_i$ with  $e_i(g_i) = \true$\@,
and, second, there is no other interaction  of higher priority for which joint participation holds. The definition of a deployable communication architecture enables us to extend the $\alpha$-th ($\alpha=1,2$) condition to the $\alpha$-th condition in distributed-enableness. The extension is by an explicit guarantee that $\sigma$ is visible by $C_i$, which can be derived from three conditions of a deployable communication architecture.

\noindent (b) We prove that $\Sigma_{dist.c} = \Sigma_c$. 
    \begin{list1}
        \item As $Com$ is a deployable communication architecture, we first prove that every distributively enabled interaction $\sigma$ is also globally enabled. Assume not, i.e., $\sigma$ is distributively-enabled but not globally enabled. This only appears (in the second condition) when another interaction $\tau$ where $\sigma \prec \tau \in \mathcal{P}$, such that $\tau$ is enabled, but $\tau$ is not visible by a component $C_i$ where $\sigma \in \Sigma_i$. This is impossible, as the definition of deployable architecture ensures that if $\sigma \prec \tau \in \mathcal{P}$,  $\sigma \in \Sigma_i$, and $\tau\in \Sigma_j$ then $C_j \informs C_i \in Com$\@, i.e., $\tau$ is visible by $C_i$. Thus $\Sigma_{dist.c} \subseteq \Sigma_c$.
        \item From (a), we have $\Sigma_c \subseteq \Sigma_{dist.c}$. Thus $\Sigma_{dist.c} = \Sigma_c$.
    \end{list1}

\noindent (c) This is the rephrasing of (a) from $A \rightarrow B$ to $\neg B \rightarrow \neg A$.
\end{proof}
\vspace{2mm}

From the above proposition (part c) we can conclude that if configuration $c$ has no distributively-enabled interaction, then $c$ is deadlocked ($c\in \mathcal{C}_{dead}$)\@.
However we are looking for an explicit guarantee for the claim that the system at configuration $c$ is never deadlocked whenever there exists one distributively-enabled interaction in $c$\@.
For our running example of memory access  in Figure~\ref{fig:VissBIP.Construction}, for example, consider the case when both {\tt C3} and {\tt D3} are
enabled  (both for allocating access to {\tt Memory3}); thus, one needs explicit assumption that the race condition will be resolved.
E.g., the run time will let {\tt Memory3} resolve the race condition and execute one of them, rather than halting permanently and
 disabling the progress. Such an assumption can be fulfilled by variants of  distributed consensus algorithms such as majority voting (MJRTY)~\cite{boyer1991mjrty}.
\vspace{2mm}
\begin{assump}[Runtime Assumption] \label{assumption2}
 For a configuration $c$ with $|\Sigma_c| >0$, the distributed controllers $\textsf{Ctrl}_i$ agree on a distributively-enabled interaction $\sigma \in \Sigma_c$ for execution.
\end{assump}
\vspace{2mm}
With the above assumption, we then define , given a system $\mathcal{S}=(C , \Sigma, \mathcal{P})$ under a communication architecture $Com$, the set of deadlock states of $\mathcal{S}$ in distributed execution to be $\mathcal{C}_{dist.dead}=\{c\}$ where no interaction is distributively-enabled at $c$. We immediately derive $\mathcal{C}_{dist.dead} = \mathcal{C}_{dead}$, as the left inclusion ($\mathcal{C}_{dist.dead} \subseteq \mathcal{C}_{dead}$) is the consequence of Proposition~\ref{prop.distributively.enableness}, and the right inclusion is trivially true.  With such an equality, given a risk configuration  $\mathcal{C}_{risk}$ and global deadlock states $\mathcal{C}_{dead}$,  we say that system $\mathit{S}$ under the distributed semantics
is \emph{distributively-safe} if there is no distributed run $c_0,\ldots,c_k$ such that $c_k\in \mathcal{C}_{dead}\cup\mathcal{C}_{risk}$;
a system that is not safe is called \emph {distributively-unsafe}\@.
Finally, we have collected all the ingredients for defining the problem of distributed priority synthesis.





\vspace{2mm}
\begin{defi} 
  Given a system $\mathcal{S} = (C , \Sigma, \mathcal{P})$  together with a deployable communication architecture $Com$, the
  set of risk configurations $\mathcal{C}_{risk} \subseteq \mathcal{C}_\sys$,
  a set of priorities $\mathcal{P}_{d+}$  is a solution to the  {\em distributed priority synthesis problem}  if the following holds:
 \begin{enumerate}
    \item  $\mathcal{P}\cup\mathcal{P}_{d+}$ is transitive and irreflexive\@.
    \item $(C , \Sigma, \mathcal{P}\cup\mathcal{P}_{d+})$ is distributively-safe\@. 
    \item  For all  $ i,j \in \{1,\ldots, m\}$ s.t. $\sigma \in \Sigma_i$,  $\tau \in \Sigma_j$,
                  if $\sigma\prec\tau\in \mathcal{P}\cup\mathcal{P}_{d+}$ then  $C_j \informs C_i \in Com$\@.
 \end{enumerate}
\end{defi}
\vspace{2mm}
The 3rd condition states that newly introduced priorities are indeed deployable.
Notice that for system $\mathcal{S}$ with a deployable communication architecture $Com$,  and any risk configurations $\mathcal{C}_{risk}$ and global deadlock states $\mathcal{C}_{dead}$,
a solution to the distributed priority synthesis problem is distributively-safe iff it is (globally) safe.
Moreover, for a fully connected communication architecture, the problem of distributed priority synthesis reduces to (global) priority synthesis.


\vspace{2mm}
\begin{theo}
Given system $\mathcal{S}=(C, \Sigma, \mathcal{P})$ under a deployable communication architecture $Com$, the problem of distributed priority synthesis is NP-complete to $|Q|+|\delta|+|\Sigma|$, where $|Q|$ and $|\delta|$ are the size of vertices and transitions in the product graph induced by $\system$\@, provided that $|C|^2 < |Q|+|\delta|+|\Sigma|$.
\end{theo}

\begin{proof}
(Sketch) First select a set of priorities (including $\mathcal{P}$) and check if they satisfy transitivity, irreflexivity, architectural constraints. Then check, in polynomial time,  if the system under this set of priorities can reach deadlock states; hardness follows from hardness of global priority synthesis. A complete proof is in the appendix.
\end{proof}
\vspace{2mm}

\begin{algorithm}[t]
\begin{scriptsize}
\DontPrintSemicolon
\SetKwRepeat{doUntil}{do}{until}
\SetKwInOut{Input}{input}\SetKwInOut{Output}{output}
\Input{Level index $i$, system $\mathcal{S} = (C=(C_1,\ldots, C_m) , \Sigma, \mathcal{P})$, communication architecture $Com$, variable set  $V_{\Sigma}$, current priority-variable assignment set $asgn$, set of deadlock states $\mathcal{C}_{dead}$ and risk states $\mathcal{C}_{risk}$}
\Output{(CONFLICT/DEADLOCK-FREE, new variable assignment)}
\Begin{
   Create $\mathcal{P}_{+}$ s.t. for all positive assignment $\underline{p}=\true$ in $asgn$, $p \in \mathcal{P}_{+}$ \;
   \nl\texttt{let} $\mathcal{P}_{tran} := \mathcal{P} \cup \mathcal{P}_{+}$\;
   \nl\doUntil{the size of $\mathcal{P}_{tran}$ does not change
   }{\lIf{$\sigma \prec \tau \in \mathcal{P}_{tran} \wedge \tau \prec \sigma' \in \mathcal{P}_{tran}$}{$\mathcal{P}_{tran} := \mathcal{P}_{tran}\cup \{\sigma\prec \sigma'\}$}
   }\nl\texttt{let} $newasgn := \emptyset$, $\Sigma_{+} := \{\sigma|\sigma\prec\sigma' \in \mathcal{P}_{+}\}\cup \{\sigma'|\sigma\prec\sigma' \in \mathcal{P}_{+}\}$\;
   \For{$\sigma\prec\tau$ in $\Sigma_{+}\times \Sigma_{+}$}{
    \lIf{$\sigma\prec\tau \in \mathcal{P}_{tran}$}{
        $newasgn := newasgn \cup \textsf{assign}(\underline{\sigma\prec\tau}, \true)$\;
    }\lElse{
        $newasgn := newasgn \cup \textsf{assign}(\underline{\sigma\prec\tau}, \false)$\;
    }
   }
   \nl\If{$\textsf{satisfy\_arch\_constraint}(\mathcal{P}_{tran}, Com)=\false \vee \textsf{satisfy\_irreflexivity}(\mathcal{P}_{tran}) = \false$}{
      \nl\textbf{return} (CONFLICT, $newasgn$)\;
   }
   \nl\texttt{let} $\mathcal{R} := \textsf{compute\_reachable}(C,\Sigma, \mathcal{P}_{tran})$\;
   \lIf{$\mathcal{R} \cap (\mathcal{C}_{dead}\cup\mathcal{C}_{risk})=\emptyset$}{

      \textbf{return} (DEADLOCK-FREE, $newasgn$)\;
    }\Else{
        \nl/* \textbf{Diagnosis-based fixing process can be inserted here} */ \;
        \nl\texttt{let} $\underline{\sigma\prec\tau} := \textsf{choose\_free\_variable}(V_{\Sigma}, newasgn)$\;
        \nl\If{$\underline{\sigma\prec\tau} \neq$ \texttt{null}}{
        \texttt{let}  $asgn1 := newasgn \cup \textsf{assign}(\underline{\sigma\prec\tau}, \true)$\;
        \texttt{let} $result := \textsf{DPS}(i+1, \mathcal{S}, Com, V_{\Sigma}, asgn1, \mathcal{C}_{dead}, \mathcal{C}_{risk})$\;
            \If{($result.1stElement$ = DEADLOCK-FREE)}{
                \textbf{return} result\;
            }\Else {
                 \texttt{let} $asgn0 := newasgn \cup \textsf{assign}(\underline{\sigma\prec \tau}, \false)$\;
                \textbf{return} $\textsf{DPS}(i+1, \mathcal{S}, Com, V_{\Sigma}, asgn0, \mathcal{C}_{dead}, \mathcal{C}_{risk})$\;
            }
        }\lElse{
            \textbf{return} (CONFLICT, $asgn$)\;
        }
    }
}
\caption{\textsf{DPS}: An algorithm for distributed priority synthesis (outline)\label{algo.prioritysyn.dpll}}
\end{scriptsize}
\end{algorithm}

\newcommand{\enc}{enc}

\section{Solving Distributed Priority Synthesis\label{sec.dps.algorithm}}


It is not difficult to derive from the NP-completeness result (Section~\ref{sec.algo.prioritysyn.distributed.execution}) a DPLL-like search algorithm (\textsf{DPS}, see Algorithm~\ref{algo.prioritysyn.dpll} for outline), where each possible priority $\sigma \prec \tau$ is represented as a Boolean variable $\underline{\sigma \prec \tau}$.
 Given $\Sigma$, let $V_{\Sigma} = \{\underline{\sigma\prec\tau}\;|\;\sigma, \tau \in \Sigma\}$ be the set of variables representing each possible priority.

    This algorithm is invoked with the empty assignment $asgn = \emptyset$\@.
    Lines 1, 2 describe the transitive closure of the current set of priorities ${\mathcal P}_+$\@.
    Then line~3 updates the assignment with $newasgn$, and
    line~4 checks if the set of derived priorities satisfies architectural constraints (using {\small $\textsf{satisfy\_arch\_constraint}$}), and is irreflexive (using {\small $\textsf{satisfy\_irreflexivity}$}).
       If not, then it returns  "conflict" in line~5. Otherwise, line~6 checks if the current set of priorities is sufficient to avoid deadlock using  reachability analysis  {\small $\textsf{compute\_reachable}$}\@.
      If successful, the current set of priorities is returned;
      otherwise, an unassigned variable $\underline{\sigma\prec\tau}$ in $V_{\Sigma}$  is chosen (using {\small $\textsf{choose\_free\_variable}$}), and,
      recursively, all possible assignments are considered  (line~8,~9)\@.
 This simple algorithm is complete as long as variables in $V_{\Sigma}$ are evaluated in a fixed order.

 Notice, however, that checking whether a risk state is reachable is expensive.
As an optimization we therefore extend the basic search algorithm above with a diagnosis-based fixing process.
In particular, whenever the system is unsafe under the current set of  priorities, the algorithm diagnoses the reason for unsafety and introduces additional priorities for preventing immediate
entry into states leading to unsafe states.
If it is possible for the current scenario to be fixed, the algorithm immediately stops and returns the fix. Otherwise, the algorithm selects a set of priorities (from reasoning the inability of fix) and uses them to guide the introduction of new priorities in \textsf{DPS}. The diagnosis-based fixing process (which is inserted in line~7 of Algorithm~\ref{algo.prioritysyn.dpll}) proceeds in two steps.

    \paragraph{Step 1: Deriving fix candidates.} Game solving is used to derive potential fix candidates represented as a set of priorities.
    In the distributed case, we  need to encode visibility constraints: they specify for each interaction $\sigma$, the set of other interactions $\Sigma_{\sigma} \subseteq \Sigma$ visible to the components executing $\sigma$ (Section~\ref{subsec.dps.core.algorithm.encoding}). With visibility constraints, our game solving process results into a \emph{nested attractor computation} (Section~\ref{subsec.dps.core.algorithm.game}). %

   \paragraph{Step 2: Fault-fixing.} We then create from fix candidates one feasible fix via solving a corresponding  SAT problem, which encodes (1) properties of priorities and (2) architectural restrictions (Section~\ref{subsec.dps.core.algorithm.SAT}).
If this propositional formula is unsatisfiable, then an  unsatisfiable  core is used to extract potentially useful  candidate  priorities.



\subsection{Game Construction\label{subsec.dps.core.algorithm.encoding}}

Symbolic encodings of interacting components form the basis of reachability checks, the diagnoses process, and  the algorithm for priority fixing (here we use $\mathcal{P}$ for $\mathcal{P}_{tran}$)\@.
In particular, symbolic encodings of system $\system = (C , \Sigma, \mathcal{P})$ use the following propositional variables:

\begin{list1}
    \item $p0$ indicates whether it is the  controller's  or the  environment's turn.
    \item  $A=\{a_{1},\dots, a_{\lceil\log_{2}|\Sigma|\rceil}\}$ for the binary encoding $\textsf{enc}(\sigma)$ of the \emph{chosen interaction} $\sigma$  (which is agreed by distributed controllers for execution, see Assumption~\ref{assumption2}). 
    \item $\bigcup_{\sigma \in \Sigma}\{\sigma\}$ are the
      variables representing interactions to encode \emph{visibility}.
      Notice that the same letter is used for an interaction and its  corresponding encoding variable.
    \item $\bigcup_{i = 1}^m Y_i $, where
      $Y_i=\{y_{i1},\dots, y_{ik}\}$  for the binary encoding $\enc(l)$ of  locations $l \in L_i$\@.
   \item $\bigcup_{i = 1}^m \bigcup_{v\in V_i} \{v\}$ are the encoding of the component variables.
\end{list1}
 Primed variables are used for encoding successor configurations and transition relations.
Visibility constraints  $\Vis^{\tau}_{\sigma} \in \{\true, \false\}$ denote the  visibility of interaction $\tau$ over another interaction $\sigma$\@.
It is computed statically: such a constraint  $\Vis^{\tau}_{\sigma}$ holds iff for $C_i, C_j \in C$ where $\tau \in \Sigma_i$ and $\sigma \in \Sigma_j $, $C_i\informs C_j \in Com$\@.

\begin{algorithm}[t]
\begin{scriptsize}
\DontPrintSemicolon
\SetKwInOut{Input}{input}\SetKwInOut{Output}{output}
\Input{System $\mathcal{S} = (C=(C_1,\ldots, C_m) , \Sigma, \mathcal{P})$, visibility constraint $\Vis_{\sigma_2}^{\sigma_1}$ where $\sigma_1,\sigma_2 \in \Sigma$}
\Output{Transition predicate $\mathcal{T}_{ctrl}$ for control and the set of deadlock states $\mathcal{C}_{dead}$}
\Begin{
     \texttt{let} predicate $\mathcal{T}_{ctrl} = \false$, $\mathcal{C}_{dead} := \true$\;
    \For{$\sigma\in \Sigma$}{
       \texttt{let} predicate $P_{\sigma} := \true$\;
     }
    \For{$\sigma\in \Sigma$}{
      \For{$i = \{1,\ldots,m\}$}{
        \nl\lIf{$\sigma \in \Sigma_i$}{
            $P_{\sigma} := P_{\sigma} \wedge \bigvee_{(l,g,\sigma,f,l')\in T_i} (\enc(l) \wedge g)$\;
         }
      }
    \nl $\mathcal{C}_{dead} := \mathcal{C}_{dead} \wedge \neg P_{\sigma}$\;
    }

    \For{$\sigma_1\in \Sigma$}{
    \nl \texttt{let} predicate $\mathcal{T}_{\sigma_1}:= p0 \wedge \neg p0' \wedge P_{\sigma_1} \wedge \textsf{enc}'(\sigma_1) \wedge \sigma_1'$\;
    \For{$\sigma_2 \in \Sigma, \sigma_2 \neq \sigma_1$}{
         \nl \lIf{$\Vis^{\sigma_2}_{\sigma_1} = \true$}{
          $\mathcal{T}_{\sigma_1} := \mathcal{T}_{\sigma_1} \wedge (P_{\sigma_2} \leftrightarrow \sigma_2')$\;
         }
         \nl\lElse{ $\mathcal{T}_{\sigma_1} := \mathcal{T}_{\sigma_1} \wedge \neg\sigma_2'$\;
         }
    }
    \For{$i = \{1,\ldots,m\}$}{
    \nl $\mathcal{T}_{\sigma_1}:= \mathcal{T}_{\sigma_1} \wedge \bigwedge_{y\in Y_i}  y \leftrightarrow y' \wedge \bigwedge_{v\in V_i} v \leftrightarrow v'$\;
    }
     \nl $\mathcal{T}_{ctrl}:= \mathcal{T}_{ctrl} \vee \mathcal{T}_{\sigma_1}$\;
    }
    \For{$\sigma_1 \prec \sigma_2 \in \mathcal{P}$}{
      \nl   $\mathcal{T}_{ctrl} := \mathcal{T}_{ctrl} \wedge
      ((\sigma_1' \wedge {\sigma_2'}) \to \neg {\textsf{enc}'(\sigma_1)})  $\;
      \nl $\mathcal{T}_{12} =\mathcal{T}_{ctrl} \wedge (\sigma'_1 \wedge \sigma'_2)$\;
      \nl $\mathcal{T}_{ctrl} := \mathcal{T}_{ctrl} \setminus \mathcal{T}_{12}$\;
      \nl $\mathcal{T}_{12,fix} := (\exists  \sigma'_1 : \mathcal{T}_{12}) \wedge (\neg \sigma'_1)$\;
      \nl $\mathcal{T}_{ctrl} := \mathcal{T}_{ctrl} \vee \mathcal{T}_{12, fix}$\;
    }

    \texttt{return} $\mathcal{T}_{ctrl}$, $\mathcal{C}_{dead}$\;
}
\caption{Generate controllable transitions and the set of deadlock states\label{algo.prioritysyn.control.transition}}
\end{scriptsize}
\end{algorithm}

\begin{algorithm}[t]
\begin{scriptsize}
\DontPrintSemicolon
\SetKwInOut{Input}{input}\SetKwInOut{Output}{output}
\Input{System $\mathcal{S} = (C=(C_1,\ldots, C_m) , \Sigma, \mathcal{P})$}
\Output{Transition predicate $\mathcal{T}_{env}$ for environment }
\Begin{
    \texttt{let} predicate $\mathcal{T}_{env} := \false$\;
    \For{$\sigma\in \Sigma$}{
        \texttt{let} predicate $T_{\sigma} := \neg p0 \wedge p0'$\;
         \For{$i = \{1,\ldots,m\}$}{
           \If{$\sigma \in \Sigma_i$}{
          \nl     $T_{\sigma} := T_{\sigma} \wedge \bigvee_{(l,g,\sigma,f,l')\in T_i}
    (\enc(l) \wedge g
    \wedge \enc'(l') \wedge  \textsf{enc}(\sigma) \wedge \textsf{enc}'(\sigma) \wedge \bigwedge_{v\in V_i} \cup_{e \in f(v)} v' \leftrightarrow e) $\;
            }
        }
        \For{$\sigma_1\in \Sigma, \sigma_1 \neq \sigma$}{
         \nl   $T_{\sigma} := T_{\sigma} \wedge \sigma_1' = \false$\;
        }
        \For{$i = \{1,\ldots,m\}$}{
           \nl \lIf{$\sigma \not\in \Sigma_i$}{
               $T_{\sigma} := T_{\sigma} \wedge \bigwedge_{y\in Y_i}  y \leftrightarrow y' \wedge \bigwedge_{v\in V_i} v \leftrightarrow v'$\;
            }
        }
        $\mathcal{T}_{env} := \mathcal{T}_{env} \vee T_{\sigma}$\;
    }
    \texttt{return} $\mathcal{T}_{env}$\;
}

\caption{Generate uncontrollable updates\label{algo.prioritysyn.environment.transition}}
\end{scriptsize}
\end{algorithm}

Algorithms~\ref{algo.prioritysyn.control.transition} and~\ref{algo.prioritysyn.environment.transition} return  symbolic transitions $\mathcal{T}_{ctrl}$  and $\mathcal{T}_{env}$  for the control players
 $\bigcup_{i=1}^{m} \textsf{Ctrl}_i$ and the player  $\textsf{Env}$ respectively,  together with the creation of a symbolic representation $\mathcal{C}_{dead}$ for the deadlock
 states of the system.
Line~1 of  algorithm~\ref{algo.prioritysyn.control.transition} computes  when an interaction $\sigma$ is enabled.
Line~2 summarizes the conditions for  deadlock, where none of the interaction is enabled. The computed deadlock condition can  be reused throughout the subsequent synthesis process, as introducing a set of priorities never introduces new deadlocks.
In line~3, $\mathcal{T}_{\sigma_1}$ constructs the actual transition, where the conjunction with $\textsf{enc}'(\sigma_1)$ indicates that $\sigma_1$ is the \emph{chosen interaction} for execution.
        $\mathcal{T}_{\sigma_1}$ is also conjoined with $\sigma_1'$ as an indication that $\sigma_1$ is enabled (and it can see itself).
Line~4 and~5 record the visibility constraint. If interaction $\sigma_2$ is visible by $\sigma_1$ ($\Vis^{\sigma_2}_{\sigma_1}=\true$), then by conjoining it  with $(P_{\sigma_2} \leftrightarrow \sigma_2')$, $\mathcal{T}_{\sigma_1}$ explicitly records the set of \emph{visible and enabled (but not chosen)} interactions. If interaction $\sigma_2$ is not visible by $\sigma_1$, then in encoding conjunct with $\neg \sigma_2'$. In this case $\sigma_2$ is \emph{treated as if it is not enabled}: if $\sigma_1$ is a bad interaction leading to the attractor of deadlock states, we cannot select $\sigma_2$ as a potential escape (i.e., we cannot create fix-candidate $\sigma_1 \prec \sigma_2$), as $\sigma_1 \prec \sigma_2$ is not supported by the visibility constraints derived by the architecture.
Line~6 keeps all variables and locations to be the same in the pre- and postcondition, as the actual update is done by the environment.
For each priority $\sigma_1 \prec \sigma_2$, lines from~8 to~12 perform transformations on the set of transitions where both $\sigma_1$ and $\sigma_2$ are enabled.
Line~8  prunes out transitions from $\mathcal{T}_{ctrl}$ where both $\sigma_1$ and $\sigma_2$ are enabled but $\sigma_1$ is chosen for execution.
Then,
 lines~9 to~12 ensure that for remaining transitions $\mathcal{T}_{12}$, they shall change the view as if $\sigma_1$ is not enabled (line~11 performs the fix). $\mathcal{T}_{ctrl}$ is updated by removing $\mathcal{T}_{12}$ and adding $\mathcal{T}_{12,fix}$.


\vspace{2mm}
\begin{prop} \label{prop.encoding.control}
Consider configuration $s$, where interaction $\sigma$ is (enabled and) chosen for execution. Given $\tau\in \Sigma$ at $s$ such that the encoding $\tau'=\true$ in Algorithm~\ref{algo.prioritysyn.control.transition}, then $\textsf{Vis}^{\tau}_{\sigma}=\true$ and interaction $\tau$ is also enabled at $s$.
\end{prop}

\begin{proof}
Assume not, i.e., there exists an interaction $\tau$ with $\tau'=\true$ in Algorithm~\ref{algo.prioritysyn.control.transition}, but either $\textsf{Vis}^{\tau}_{\sigma}= \false$ or $\tau$ is not enabled.
\begin{list1}
    \item If $\textsf{Vis}^{\tau}_{\sigma}= \false$, then line~5 explicitly sets $\tau'$ to \false; if $\tau = \sigma$ then Assumption~1 ensures that $\textsf{Vis}^{\tau}_{\sigma}= \true$. Both lead to contradiction.
    \item If $\tau$ is not enabled, based on the definition, there are two reasons.
        \begin{list1}
            \item There exists another interaction $\kappa \neq \sigma$ such that $\kappa$ is enabled at $s$ and priority $\tau \prec \kappa$ exists. In this case, then line~9 to~12 ensures that $\tau'=\false$. Contradiction.
            \item $\tau$ is not enabled as it does not satisfy the precondition. For this line~4 ensures that if $\tau$ is not enabled, $\tau'$ is set to \false. Contradiction.
        \end{list1}
\end{list1}
\end{proof}
\vspace{2mm}

\begin{prop}\label{prop.encoding.deadlock}
 $\mathcal{C}_{dead}$  as returned by algorithm~\ref{algo.prioritysyn.control.transition} encodes the set of deadlock states of the input system $\mathcal{S}$\@.
\end{prop}

\begin{proof} We first recap that using priorities never introduces new deadlocks, as (1) $\sigma \prec \tau$ only blocks $\sigma$ when $\tau$ is enabled, and (2) for $\mathcal{P}$, its defined relation is transitive and irreflexive (so we never have cases like $\sigma_1\prec \sigma_2 \prec \sigma_3 \prec \ldots \prec \sigma_1$, which creates $\sigma_1 \prec \sigma_1$, violating irreflexive rules). 
\begin{list1}
    \item The set of deadlock states for distributed execution, based on Assumption~1 and~2, amounts to the set of global deadlock states, where each interaction is not enabled. Based on the definition, situations where an interaction $\sigma$ is not enabled can also occur when its guard condition holds, but there exists another interaction $\tau$ such that (1) the guard-condition of $\tau$ holds on all components, and (2) $\sigma \prec \tau$ exists.
         \begin{list1}
            \item If $\tau$ is not blocked by another interaction, then $\tau$ is enabled for execution, so such a case never constitutes new deadlock states.
            \item Otherwise, we can continue the chain process and find an interaction $\kappa$ (this chain never repeats back to $\tau$, based on above descriptions on properties of priorities) whose guard-condition holds and is not blocked. Then no new deadlock is introduced.
         \end{list1}
         Therefore, deadlock only appears in the case where for each interaction, its guard-condition does not hold. This condition is computed  by the loop over each interaction with line~2.
\end{list1}
\end{proof}
\vspace{2mm}

In Algorithm~\ref{algo.prioritysyn.environment.transition}, the environment updates the configuration using interaction $\sigma$ based on the indicator $\textsf{enc}(\sigma)$. Its freedom of choice in variable updates is listed in line~1 (i.e., $\cup_{e \in f(v)} v' \leftrightarrow e$). Line~2 explicitly sets all interactions $\sigma_1$ not cosen for execution to be false, and line~3 sets all components not participated in $\sigma$ to be stuttered.

Finally,  Figure~\ref{fig:vissbip.nested.computation} exemplifies an encoding for control (represented by a  circle); the current system configuration is assumed to be
 $c_1$, and it is  assumed that both $\sigma_1$ and $\sigma_2$ can be executed, but $\Vis^{\sigma_2}_{\sigma_1} = \Vis^{\sigma_1}_{\sigma_2} = \false$\@.


\newcommand{\Occ}{\ensuremath{\textrm{Occ}}}
\newcommand{\Inf}{\ensuremath{\textrm{Inf}}}
\newcommand{\attr}{\ensuremath{\textsf{attr}}}
\newcommand{\N}{\mathbf{N}}


\begin{algorithm}[t]
\begin{scriptsize}
\DontPrintSemicolon
\SetKwInOut{Input}{input}\SetKwInOut{Output}{output}
\Input{Initial state $c_0$, risk states $\mathcal{C}_{risk}$, deadlock states $\mathcal{C}_{dead}$, set of reachable states $\reach(\{c_0\})$ and symbolic transitions $\mathcal{T}_{ctrl}$, $\mathcal{T}_{env}$ from Algorithm~\ref{algo.prioritysyn.control.transition} and~\ref{algo.prioritysyn.environment.transition} }
\Output{(1) Nested risk attractor $\textsf{NestAttr}_{env}(\mathcal{C}_{risk}\cup \mathcal{C}_{dead})$ and (2) $\mathcal{T}_{f} \subseteq \mathcal{T}_{ctrl}$, which is the set of control transitions starting outside $\textsf{NestAttr}_{env}(\mathcal{C}_{dead}\cup\mathcal{C}_{risk})$ but entering $\textsf{NestAttr}_{env}(\mathcal{C}_{risk}\cup \mathcal{C}_{dead})$.}
\Begin{
    \tcp{Create architectural non-visibility predicate}
    \nl\textbf{let} $\textsf{Esc} := \false$\;
    \For{$\sigma_i \in \Sigma$}{
        \nl\textbf{let} $\textsf{Esc}_{\sigma_i} := \textsf{enc}'(\sigma_i)$\;
         \nl \lFor{$\sigma_j \in \Sigma, \sigma_j \neq \sigma_i$}{
            $\textsf{Esc}_{\sigma_i} := \textsf{Esc}_{\sigma_i} \wedge \neg\sigma_j'$\;
         }
         $\textsf{Esc} := \textsf{Esc} \vee (\textsf{Esc}_{\sigma_i} \wedge \sigma_i')$\;
        }
    \tcp{Part A: Prune unreachable transitions and bad states}
    $\mathcal{T}_{ctrl} := \mathcal{T}_{ctrl} \wedge \reach(\{c_0\})$, $\mathcal{T}_{env} := \mathcal{T}_{ctrl} \wedge \reach(\{c_0\})$\;
    $\mathcal{C}_{dead} := \mathcal{C}_{dead} \wedge \reach(\{c_0\})$, $\mathcal{C}_{risk} := \mathcal{C}_{risk} \wedge \reach(\{c_0\})$\;
    \;
    \tcp{Part B: Solve nested-safety game}
    \textbf{let} $\textsf{NestedAttr}_{pre} := \mathcal{C}_{dead} \vee \mathcal{C}_{risk}$,  $\textsf{NestedAttr}_{post} := \false$\;
    \nl\While{$\true$}{
    \textbf{let} $\Attr_{pre} := \textsf{NestedAttr}_{pre}$,  $\Attr_{post} := \false$\;
    \;
    \tcp{B.1 Compute risk attractor}
    \nl \While{\true}{
    \tcp{add environment configurations}
     $\Attr_{post,env} := \exists\Xi': (\mathcal{T}_{env} \wedge \texttt{SUBS}((\exists \Xi': \Attr_{pre}), \Xi, \Xi'))$\;
    \tcp{add system configurations}
    \textbf{let} $\texttt{PointTo} := \exists\Xi': (\mathcal{T}_{ctrl} \wedge \texttt{SUBS}((\exists \Xi': \Attr_{pre}), \Xi, \Xi'))$\;
     \textbf{let} $\texttt{Escape} := \exists\Xi': (\mathcal{T}_{ctrl} \wedge \texttt{SUBS}((\exists \Xi': \neg \Attr_{pre}), \Xi, \Xi'))$\;
     $\Attr_{post,ctrl} := \texttt{PointTo} \setminus \texttt{Escape}$\;
     $\Attr_{post} := \Attr_{pre} \vee \Attr_{post,env} \vee \Attr_{post,ctrl} $\tcp*[r]{Union the result}
     \lIf{$\Attr_{pre} \leftrightarrow \Attr_{post}$}{
        \texttt{break}\tcp*[r]{Break when the image saturates}
     }
     \lElse{
     $\Attr_{pre} := \Attr_{post}$\;
     }
    }
    \;
    \tcp{B.2 Generate transitions with source in $\neg \Attr_{pre}$ and destination in $\Attr_{pre}$}
     \nl $\texttt{PointTo} := \mathcal{T}_{ctrl} \wedge \texttt{SUBS}((\exists \Xi': \Attr_{pre}), \Xi, \Xi'))$\;
    \nl $\texttt{OutsideAttr} := \neg \Attr_{pre} \wedge (\exists \Xi': \mathcal{T}_{ctrl})$\;
    \nl $\mathcal{T} := \texttt{PointTo} \wedge \texttt{OutsideAttr}$\;
    \;
    \tcp{B.3 Add the source vertex of B.2 to $\textsf{NestedAttr}_{post}$, if it can not see another interaction for escape}
     \nl $\textsf{newBadStates} := \exists \Xi':(\mathcal{T} \wedge \textsf{Esc})$\;
     \nl $\textsf{NestedAttr}_{post} := \Attr_{pre} \vee \textsf{newBadStates}$\;
    \tcp{B.4 Condition for breaking the loop}
     \lIf{$\textsf{NestedAttr}_{pre} \leftrightarrow \textsf{NestedAttr}_{post}$}{
        \texttt{break}\tcp*[r]{Break when the image saturates}
     }\lElse{
     $\textsf{NestedAttr}_{pre} := \textsf{NestedAttr}_{post}$\;
     }
    }
    \;
    \tcp{Part C: extract $\mathcal{T}_{f}$ }
    \nl $\texttt{PointToNested} := \mathcal{T}_{ctrl} \wedge \texttt{SUBS}((\exists \Xi': \textsf{NestedAttr}_{pre}), \Xi, \Xi'))$\;
    \nl $\texttt{OutsideNestedAttr} := \neg \textsf{NestedAttr}_{pre} \wedge (\exists \Xi': \mathcal{T}_{ctrl})$\;
    \nl $\mathcal{T}_{f} := \texttt{PointToNested} \wedge \texttt{OutsideNestedAttr}$\;
    \;
    \texttt{return} $\textsf{NestAttr}_{env}(\mathcal{C}_{dead}\cup\mathcal{C}_{risk}):=\textsf{NestedAttr}_{pre}$, $\mathcal{T}_{f}$\;
}

\caption{Nested-risk-attractor computation\label{algo.prioritysyn.nested.attractor}}
\end{scriptsize}
\end{algorithm}

\subsection{Fixing Algorithm:  Game Solving with Nested Attractor Computation\label{subsec.dps.core.algorithm.game}}

The first step of fixing is to compute the \emph{nested-risk-attractor} from the set of bad states $\mathcal{C}_{risk}\cup \mathcal{C}_{dead}$. Let $V_{ctrl}$ ($\mathcal{T}_{ctrl}$) and $V_{env}$ ($\mathcal{T}_{env}$) be the set of control and environment states (transitions) in the encoded game. Let \emph{risk-attractor} $\Attr_{env}(X):=\bigcup_{k\in\N} \attr^k_{env}(X)$, where
\[
 \attr_{env}(X) := X \cup \{ v\in V_{env} \mid v\mathcal{T}_{env} \cap X \neq \emptyset \} \cup \{ v\in V_{ctrl} \mid \emptyset \neq v\mathcal{T}_{ctrl} \subseteq X \},
\]
i.e., $\attr_{env}(X)$ extends state sets $X$ by all those states from which either environment can move to $X$ within one step or control
cannot prevent to move within the next step. ($v\mathcal{T}_{env}$ denotes the set of environment successors of $v$, and $v\mathcal{T}_{ctrl}$ denotes the set of control successors of $v$.)
Then $\Attr_{env}(\mathcal{C}_{risk}\cup \mathcal{C}_{dead}) := \bigcup_{k\in\N} \attr^k_{env}(\mathcal{C}_{risk}\cup \mathcal{C}_{dead})$ contains all nodes from which environment can force any play to visit the set $\mathcal{C}_{risk}\cup \mathcal{C}_{dead}$.

\begin{figure}
\centering
 \includegraphics[width=0.6\columnwidth]{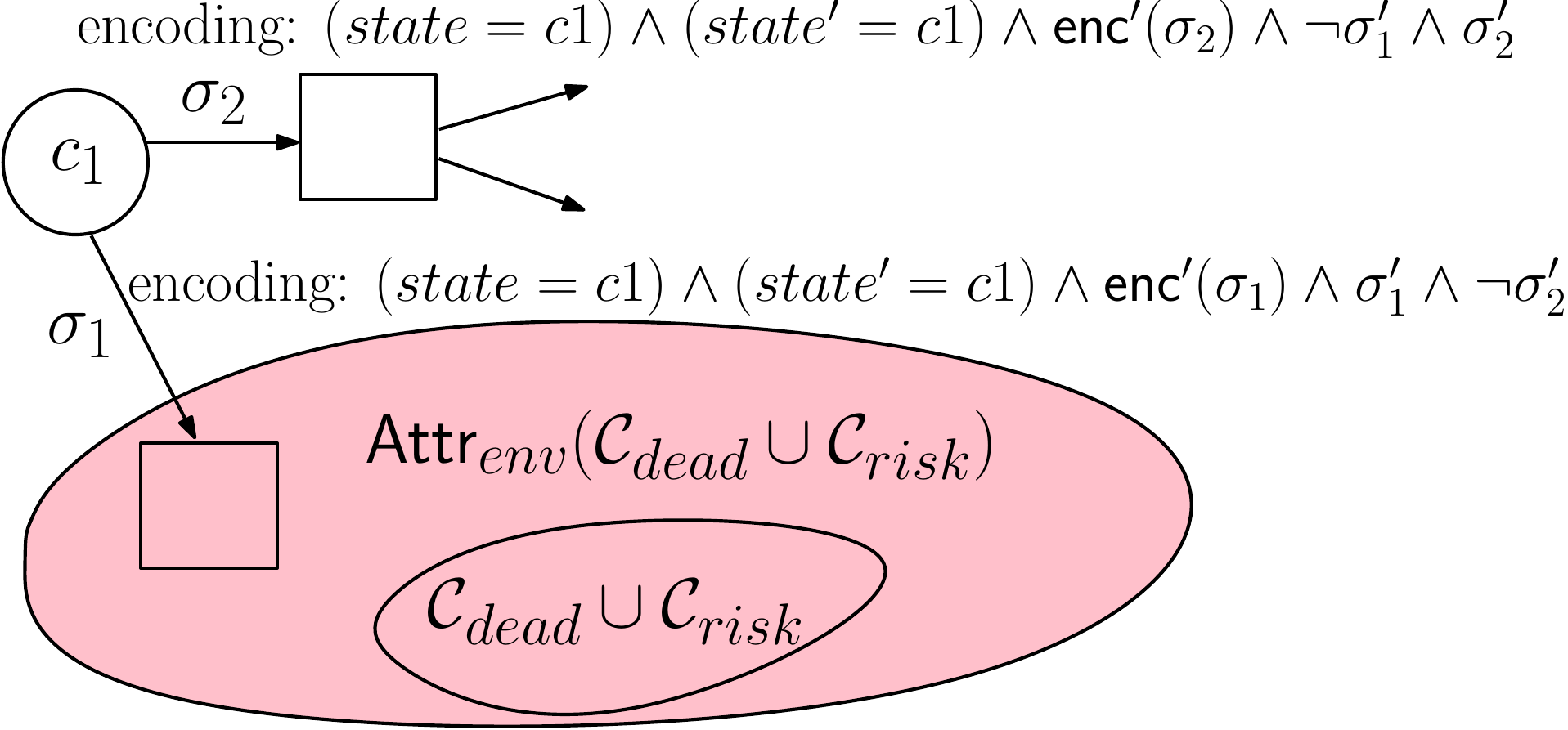}
  \caption{Intermediate nested computation: scenario when System $\mathcal{S}$ is in configuration $c_1$, which is outside the attractor but $\Vis^{\sigma_2}_{\sigma_1}=\Vis^{\sigma_1}_{\sigma_2}=\false$.}
 \label{fig:vissbip.nested.computation}
\end{figure}

Nevertheless, nodes outside the risk-attractor are not necessarily safe due to visibility constraints. Figure~\ref{fig:vissbip.nested.computation} illustrates such a concept. Configuration $c_1$ is a control location, and it is outside the attractor: although it has an edge $\sigma_1$ which points to the risk-attractor, it has another edge $\sigma_2$, which does not lead to the attractor. We call positions like $c_1$ as \emph{error points}. Admittedly, applying priority $\sigma_1 \prec \sigma_2$ at $c_1$ is sufficient to avoid entering the attractor. However, as $\Vis^{\sigma_2}_{\sigma_1} = \false$, then for components who try to execute $\sigma_1$, they are unaware of the enableness of $\sigma_2$. So $\sigma_1$ can be executed freely. Therefore, we should add $c_1$ explicitly to the (already saturated) attractor, and recompute the attractor due to the inclusion of new vertices.
This leads to an extended computation of the risk-attractor (i.e., nested-risk-attractor).
\vspace{2mm}
\begin{defi}
The \emph{nested-risk-attractor} $\textsf{NestAttr}_{env}(\mathcal{C}_{risk}\cup \mathcal{C}_{dead})$  is the smallest superset of $\Attr_{env}(\mathcal{C}_{risk}\cup \mathcal{C}_{dead})$ such that the following holds.
\begin{enumerate}
\item For state $c \not\in \textsf{NestAttr}_{env}(\mathcal{C}_{risk}\cup \mathcal{C}_{dead})$, where these exists a (bad-entering) transition $t\in \mathcal{T}_{ctrl}$
with source $c$ and target $c' \in \textsf{NestAttr}_{env}(\mathcal{C}_{risk}\cup \mathcal{C}_{dead})$:
    \begin{list1}
    \item \emph{(Good control state shall have one escape)} there exists another transition $t'\in \mathcal{T}_{ctrl}$ such that its source is $c$ but its destination $c'' \not\in \textsf{NestAttr}_{env}(\mathcal{C}_{risk}\cup \mathcal{C}_{dead})$.
    \item\emph{ (Bad-entering transition shall have another visible candidate)} for every bad-entering transition $t$ of $c$, in the encoding let $\sigma$ be the chosen interaction for execution ($\textsf{enc}'(\sigma) = \true$). Then there exists another interaction $\tau$ such that, in the encoding, $\tau'=\true$.
    \end{list1}
\item \emph{(Add if environment can enter)} If $v\in V_{env}$, and $v\mathcal{T}_{env} \cap \textsf{NestAttr}_{env}(\mathcal{C}_{risk}\cup \mathcal{C}_{dead})  \neq \emptyset$, then $v \in \textsf{NestAttr}_{env}(\mathcal{C}_{risk}\cup \mathcal{C}_{dead})$.
\end{enumerate}
\end{defi}
\vspace{2mm}

Algorithm~\ref{algo.prioritysyn.nested.attractor} uses a nested fixpoint for computing a symbolic representation of  a nested risk attractor.
The notation $\exists \Xi$ ($\exists \Xi'$) is used to represent existential quantification over all umprimed (primed) variables used in the system encoding.
Moreover,  we use the operator $\texttt{SUBS}(X, \Xi, \Xi')$, as available in many BDD packages, for variable swap (substitution) from unprimed to primed variables in $X$\@.
For preparation (line~1 to~3), we first create a predicate, which explicitly records when an interaction $\sigma_i$ is enabled and chosen (i.e., $\sigma_i' = \true$ and $\textsf{enc}'(\sigma_i) = \true$). For every other interaction $\sigma_j$, the variable $\sigma_j'$ is evaluated to $\false$ in BDD (i.e., either it is disabled or not visible by $\sigma_i$, following Algorithm~\ref{algo.prioritysyn.control.transition}, line~4 and~5).

The nested computation consists of two while loops (line~4,~5):
        the inner while loop B.1 computes the familiar risk attractor, and
         B.2 computes the set of transitions $\mathcal{T}$ whose source is outside the attractor but the destination is inside the attractor. Notice that for every source vertex $c$ of a transition in $\mathcal{T}$: (1) It has chosen an interaction $\sigma \in \Sigma$ to execute, but it is a bad choice. (2) There exists another choice $\tau$ whose destination is outside the attractor (otherwise, $c$ shall be in the attractor).
            However, such $\tau$ may not be visible by $\sigma$. Therefore, $\exists \Xi':(\mathcal{T} \wedge \textsf{Esc})$ creates those states without any \emph{visible escape}, i.e., without any other visible and enabled interactions under the local view of the chosen interaction. These states form the set of new bad states $\textsf{newBadStates}$ due to architectural limitations.

A visible escape is not necessarily a  "true escape" as illustrated in Figure~\ref{fig:vissbip.locate.fix}.
It is possible that for state $c_2$, for $g$ its visible escape is $a$, while for $a$ its visible escape is $g$. Therefore, it only suggests candidates of fixing, and in these cases, a feasible fix is derived in  a SAT resolution step (Section~\ref{subsec.dps.core.algorithm.SAT}).
Finally,  Part~C of the algorithm extracts  $\mathcal{T}_{f}$ (similar to extracting $\mathcal{T}$ in B.2)\@.

Consider again the situation depicted in Figure~\ref{fig:vissbip.nested.computation}. In Algorithm~\ref{algo.prioritysyn.nested.attractor}, after the attractor is computed, lines~6-8 extract the symbolic transition $(state = c1) \wedge(state' = c1) \wedge  \textsf{enc}'(\sigma_1) \wedge  \sigma'_1 \wedge \neg \sigma'_2$.
Then by a conjunction with $\textsf{Esc}$ (from line~1 to~3) and performing quantifier elimination over primed variables, one recognizes that $c1$ shall
 be added to \textsf{newBadStates}; the algorithm continues with the next round of nested computation.

  Algorithm~\ref{algo.prioritysyn.nested.attractor} terminates, since  the number of states that can be added to $\textsf{Attr}_{post}$ (in the inner-loop) and $\textsf{NestedAttr}_{post}$ (in  the outer-loop) is finite.
The following proposition is used to  detect  the infeasibility of  distributed priority synthesis problems. 

\vspace{2mm}
\begin{prop}\label{prop.encoding.fast.diagnose}
Assume during the base-level execution of Algorithm~\ref{algo.prioritysyn.dpll} where $asgn=\emptyset$\@.
If the encoding of the initial state is contained in $\textsf{NestAttr}_{env}(\mathcal{C}_{risk}\cup \mathcal{C}_{dead})$, then the
distributed priority synthesis problem for $\mathcal{S}$ with  $\mathcal{C}_{risk}$  is infeasible.
\end{prop}

\begin{proof}
In Algorithm~\ref{algo.prioritysyn.dpll}, when the fixing process is invoked at the base level where $asgn=\emptyset$, $\mathcal{P}_{tran} = \mathcal{P}$. Assume after the execution of the nested-risk-attractor (Algorithm~\ref{algo.prioritysyn.nested.attractor}), the symbolic encoding of the initial state $c_0$ (which is a control state) is in $\textsf{NestAttr}_{env}(\mathcal{C}_{risk}\cup \mathcal{C}_{dead})$.
Then based on Algorithm~\ref{algo.prioritysyn.nested.attractor}, the encoded state of $c_0$ is added to $\textsf{NestAttr}$ because
\begin{list1}
    \item either all of its edges enter the previously computed $\textsf{NestAttr}$ (in this case, no priority can help to block the entry),
    \item or it has a transition which enters the previously computed $\textsf{NestAttr}$ with interaction $\sigma$ but has no visible escape $\tau$, i.e., in the encoding of the transition, $\textsf{enc}'(\sigma)=\true$ and for all $\tau\in \Sigma, \tau \neq \sigma$, we have encoding $\tau'=\false$. From the encoding how $\tau'$ is set to false, we know that for such a transition, for any fix of the form   $\sigma \prec \tau$, it is either not supported by the architecture (see Algorithm~\ref{algo.prioritysyn.control.transition} for encoding, line~5), or $\tau$ is not enabled at $c_0$ (Algorithm~\ref{algo.prioritysyn.control.transition}, line~4). Therefore, in the distributed execution, executing $\sigma$ at $c_0$ can not be blocked by the use of priority.
\end{list1}
Overall, this leads to the entry of the previously computed $\textsf{NestAttr}$. Continuing the process we can conclude that $\mathcal{C}_{risk}\cup \mathcal{C}_{dead}$ can be reached, and no priority can assist to escape from entering. Consider when analysis is done at the base level where $\mathcal{P}_{tran} = \mathcal{P}$, then there exists no $\mathcal{P}_{d+}$ as a solution of the distributed priority synthesis problem.

The number of required steps of entering is no larger than $outer\times inner$ steps, where $outer$ is the number of iterations for the outer-while-loop, and $inner$ is the maximum number of iterations for all inner-while-loop execution.
\end{proof}
\vspace{2mm}

\subsection{Fixing Algorithm: SAT Problem Extraction and Conflict Resolution\label{subsec.dps.core.algorithm.SAT}}

\begin{figure}
\centering
 \includegraphics[width=0.6\columnwidth]{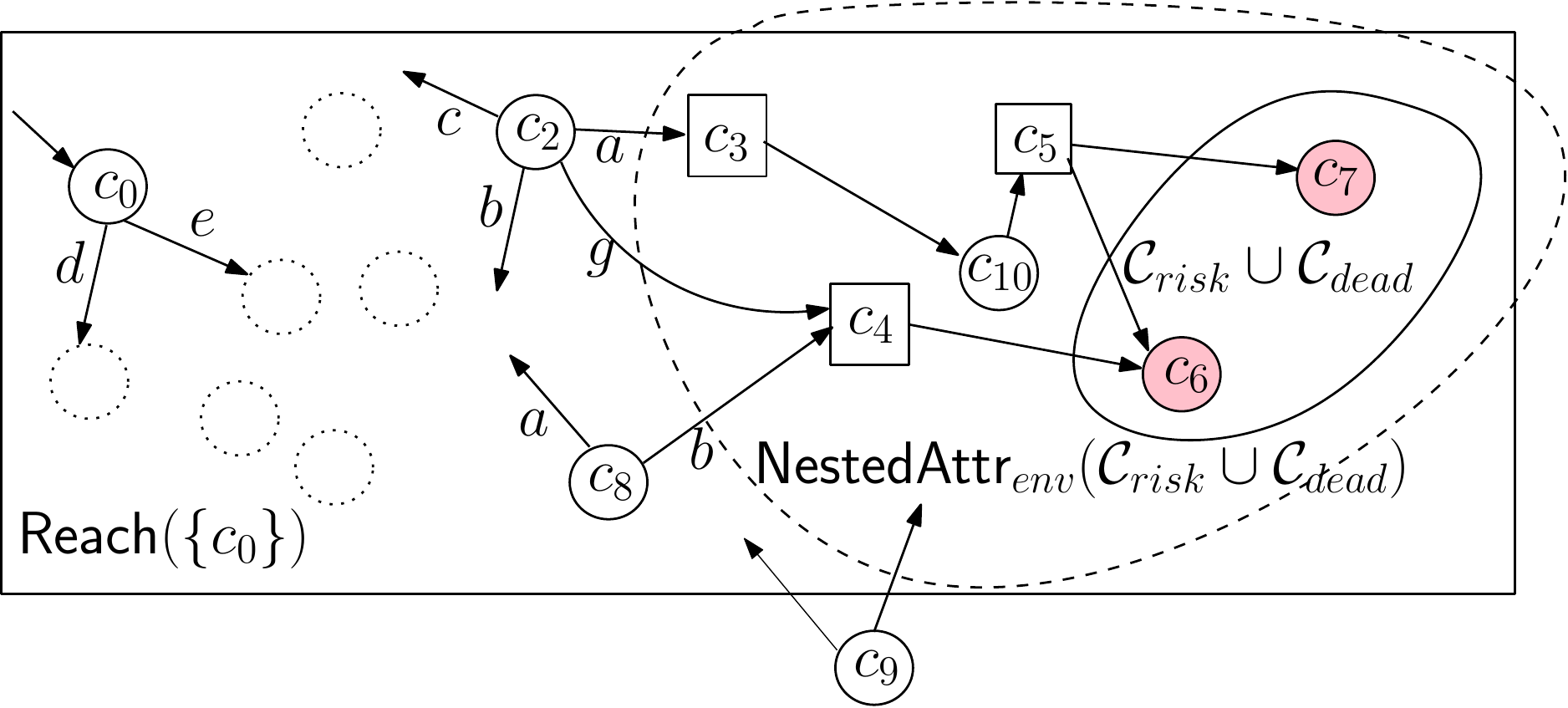}
  \caption{Locating fix candidates outside from the nested-risk-attractor.}
 \label{fig:vissbip.locate.fix}
\end{figure}

 The return value
$\mathcal{T}_{f}$
of Algorithm~\ref{algo.prioritysyn.nested.attractor}
contains not only the risk interactions but also all
possible interactions which are visible and enabled (see Algorithm~\ref{algo.prioritysyn.control.transition} for encoding, Proposition~\ref{prop.encoding.control} for result).
Consider, for example, the situation depicted  in Figure~\ref{fig:vissbip.locate.fix} and assume that  $\Vis^{c}_{a}$, $\Vis^{b}_{a}$, $\Vis^{c}_{b}$, $\Vis^{a}_{g}$,  and $\Vis^{a}_{b}$ are the only visibility constraints which hold \true.
 If $\mathcal{T}_{f}$ returns three transitions, one may extract fix candidates from each of these transitions in the following way.
\begin{list1}
    \item On $c_2$, $a$ enters the nested-risk-attractor, while $b,c$ are also visible from $a$; one obtains the  candidates $\{a \prec b, a \prec c\}$.
    \item On $c_2$, $g$ enters the nested-risk-attractor, while $a$ is also visible from $g$; one obtains the  candidate $\{g \prec a\}$.
    \item On $c_8$, $b$ enters the nested-risk-attractor, while $a$ is also visible; one obtains the candidate $\{b \prec a\}$.
\end{list1}
Using these candidates, one can perform \emph{conflict resolution} and
generate a set of new  priorities for preventing entry into the nested-risk-attractor region.
For example, $\{a\prec c, g \prec a, b \prec a\}$ is such a set of priorities for ensuring  the safety condition.
Notice also  that the set $\{a\prec b, g \prec b,
b \prec a\}$ is circular, and therefore not a valid set of priorities.

In our implementation, conflict resolution is performed using SAT solvers.
Priorities $\sigma_1 \prec \sigma_2$ are  presented as a Boolean variable $\underline{\sigma_1
  \prec \sigma_2}$\@.
 If the generated SAT problem is satisfiable, for all variables
$\underline{\sigma_1 \prec \sigma_2}$ which is evaluated to $\true$,
we add priority $\sigma_1 \prec \sigma_2$ to the resulting introduced priority set $\mathcal{P}_{d+}$.
The constraints below correspond to the ones for  global priority synthesis framework~\cite{cheng:algo.priority.syn:2011}\@.
\begin{enumerate}
    \item \emph{(Priority candidates)} For each edge $t \in \mathcal{T}_{f}$ which enters the risk attractor using $\sigma$ and having $ \sigma_1, \ldots, \sigma_e$ visible escapes (excluding $\sigma$), create clause $(\bigvee_{i=1}^e \underline{\sigma \prec \sigma_{i}})$.\footnote{In implementation, Algorithm~\ref{algo.prioritysyn.nested.attractor} works symbolically on BDDs and proceeds on \emph{cubes} of the risk-edges (a cube contains a set of states having the same enabled interactions and the same risk interaction), hence it avoids enumerating edges state-by-state.}
    \item \emph{(Existing priorities)} For each priority $\sigma \prec \tau \in \mathcal{P}$, create
        clause $(\underline{\sigma \prec \tau})$.
    \item \emph{(Irreflexive)} For each interaction $\sigma$ used in (1) and (2), create clause
            $(\neg \underline{\sigma \prec \sigma})$.
    \item \emph{(Transitivity)} For any  $\sigma_1, \sigma_2, \sigma_3$  used above,  create a clause
            $((\underline{\sigma_1 \prec \sigma_2} \wedge \underline{\sigma_2 \prec \sigma_3})\Rightarrow \underline{\sigma_1 \prec \sigma_3})$.
\end{enumerate}
Clauses for architectural constraints also need to be added in the case of distributed priority synthesis.
For example, if $\sigma_1 \prec \sigma_2$ and $\sigma_2 \prec \sigma_3$ then due to transitivity we shall include priority $\sigma_1 \prec \sigma_3$. But if $\Vis^{\sigma_3}_{\sigma_1} = \false$, then $\sigma_1 \prec \sigma_3$ is not supported by communication.  In the above example, as $\Vis^{c}_{b} = \true$, $\{a\prec c, g \prec a, b \prec a\}$ is a legal set of priority fix satisfying the architecture (because the inferred priority $b \prec c$ is supported). Therefore, we introduce the following constraints.
\begin{list1}
\item \emph{(Architectural Constraint)} Given $\sigma_1,\sigma_2\in \Sigma$, if $\Vis^{\sigma_2}_{\sigma_1} = \false$, then $\underline{\sigma_1 \prec \sigma_2}$ is evaluated to $\false$.
\item \emph{(Communication Constraint)} Given $\sigma_1,\sigma_2\in \Sigma$, if $\Vis^{\sigma_2}_{\sigma_1} = \false$, for any interaction $\sigma_3\in \Sigma$, if $\Vis^{\sigma_3}_{\sigma_1} = \Vis^{\sigma_2}_{\sigma_3} = \true$, at most one of $\underline{\sigma_1 \prec \sigma_3}$ or $\underline{\sigma_3 \prec \sigma_2}$ is evaluated to $\true$.
\end{list1}

\noindent
A correctness argument of this fixing process  can be found  in the appendix.

\section{Implementation\label{sec.dps.algo.extension}}

Our algorithm for solving the distributed priority synthesis problem has been implemented in Java on top of the open-source workbench VissBIP\footnote{Available from \url{http://www.fortiss.org/formal-methods}\@.}
for graphically editing and visualizing systems of interacting components. The synthesis engine itself is based on  the
JDD package for binary decision diagrams, and the SAT4J  propositional satisfiability solver.
In addition, we implemented a number of extensions and optimizations (e.g., Proposition~\ref{prop.encoding.fast.diagnose}) to the core algorithm in Section~\ref{sec.dps.algorithm}; for lack of space details needed to be omitted\@.

First, we also use the result of the unsatisfiable core during the fix process to guide the assignment of variables (where each represents a priority) in the \textsf{DPS} algorithm. E.g., if the fix does not succeed as both $\sigma \prec \tau$ and $\tau \prec \sigma$ are used, the engine then introduces $\underline{\sigma \prec \tau}$. Then in the next diagnosis process, the engine can not propose a fix of the form $\tau \prec \sigma$ (as to give such a fix by the engine, it requires that when $\tau$ and $\sigma$ are enabled while $\tau$ is chosen for execution, $\sigma$ is also enabled; the enableness of $\sigma$ contradicts $\sigma \prec \tau$). 

Second, we are over-approximating the nested risk attractor by parsimoniously adding all source states in $\mathcal{T}_f$, as returned
from Algorithm~\ref{algo.prioritysyn.nested.attractor}, to the nested-risk-attractor before recomputing; thereby increasing
chances of creating a new  $\mathcal{T}_f$ where conflicts can be resolved.

Lastly, whenever possible the implementation tries to synthesize a local controllers without any state information. If such a diagnosis-fixing fails, the algorithm can also perform a model transformation of the interacting components which is equivalent to transmitting state information in the communication. Recall that the symmetric communication architecture in Figure~\ref{table:VissBIP.result} requires communicating not only of the intended next moves but also of the current source locations.
In order to minimize the amount  of state information that is required to communicate, we lazily extract refinement candidates  from (minimal) unsatisfiable cores of failed runs of the propositional solver, and correspondingly refine the alphabet by including new state information. Alternatively, a fully refined model transformation can eagerly be computed in VissBIP\@.




\section{Evaluation\label{sec.dps.evaluation}}

We validate our algorithm  using a collection of benchmarking models
including  memory access problem, power allocation assurance, and working protection in industrial automation; some of these case studies are extracted from industrial case studies.  Table~\ref{tab:exp} summarizes the results obtained on  an Intel Machine with 3.4\,GHz CPU and 8\,GB RAM\@. Besides  runtime we also list the algorithmic extensions and optimizations described in Section~\ref{sec.dps.algo.extension}\@.

The experiments 1.1 through 1.16 in Table~\ref{tab:exp} refer to variations of the multiprocessor scheduling problem with increasing number of processors and memory banks. Depending on the communication architectures the engine uses refinement or extracts the UNSAT core to find a solution.

Experiments 2.1 and 2.2 refer to a multi-robot scenario with possible moves in a predefined arena, and the goal is to avoid collision by staying within a predefined protection cap.
The communication architecture is restricted in that the $i$-th robot can only notify the $((i+1)\% n)$-th\@.
\comment{
   \item \textbf{(Automation safety)} In an arena, every robot can make a set of possible movements to four directions. There exists more than one robot in the same arena. The risk in the current configuration is that all the robots may collapse in the same position. Our purpose is to form a rule-based protection cap to ensure that all the robots appearing simultaneously in the same location will never happen. Here robots are not fully informed, i.e., we explicitly define the communication structure as follows: $Robot[i]$ can notify $Robot[(i+1) \mod n]$.
}

In experiments 3.1 through 3.6 we investigate the classical dining philosopher problem using various communication architectures.  If the communication is clockwise, then the engine fails to synthesize priorities\footnote{Precisely, in our model, we allow each philosopher to pass his intention over his left fork to the philosopher of his left. The engine uses Proposition~4 and diagnoses that it is impossible to synthesize priorities, as the initial state is within the nested-risk-attractor. }. If the communication is counter-clockwise (i.e., a philosopher can notify its intention to his right philosopher), then the engine is also able to synthesize distributed priorities (for $n$ philosophers, $n$ rules suffice). Compared to our previous priority synthesis technique, as in distributed priority synthesis we need to separate visibility and enabled interactions, the required time for synthesis is longer.

Experiment 4 is based on a case study for increasing the reliability of data processing units (DPUs) by using multiple data sampling. The mismatch between the calculated results from different devices may yield deadlocks. The deadlocks can be avoided with the synthesized priorities from VissBIP without modifying local behaviors.

Finally, in experiment 5, we are synthesizing a decentralized controller for the Dala robot~\cite{joser11}, which is composed of 20 different  components. A hand-coded version of the  control indeed did not rule out deadlocks.
Without any further communication constraints between the components, VissBIP locates the deadlocks and synthesizes additional priorities to avoid them.

\begin{table}[t]
\caption{Experimental results on distributed priority synthesis}\label{tab:exp}
\centering
\begin{scriptsize}
\begin{threeparttable}
\begin{tabular}{|l|l | c | c |c| l|}\hline
Index & Testcase and communication architecture & Components & Interactions & Time (seconds) & Remark\\\hline
1.1 & 4 CPUs with broadcast A & 8 &24 & 0.17 & x \\\hline
1.2 & 4 CPUs with local A, D & 8 &24 & 0.25 & A \\\hline
1.3 & 4 CPUs with local communication & 8 &24 & 1.66 & R \\\hline
1.4 & 6 CPUs with broadcast A & 12 &36 & 1.46 & RP-2 \\ \hline
1.5 & 6 CPUs with broadcast A, F & 12 &36 & 0.26 & x \\\hline
1.6 & 6 CPUs with broadcast A, D, F & 12 &36 & 1.50 & A \\\hline
1.7 & 6 CPUs with local communication  & 12 &36 & -  & fail \\\hline
1.8 & 8 CPUs with broadcast A & 16 &48 & 8.05 & RP-2 \\\hline
1.9 & 8 CPUs with broadcast A, H & 16 &48 & 1.30 & x \\\hline
1.10 & 8 CPUs with broadcast A, D, H & 16 & 48 & 1.80 & x \\\hline
1.11 & 8 CPUs with broadcast A, B, G, H & 16 & 48 & 3.88 & RP-2 \\\hline
1.12 & 8 CPUs with local communication & 16 & 48 & 42.80 & R \\\hline
1.13 & 10 CPUs with broadcast A & 20 & 60 & 135.03 & RP-2 \\\hline
1.14 & 10 CPUs with broadcast A, J & 20 & 60 & 47.89 & RP-2 \\\hline
1.15 & 10 CPUs with broadcast A, E, F, J  & 20 & 60 & 57.85 & RP-2 \\\hline
1.16 & 10 CPUs with local communication A, B, E, F, I, J  & 20 & 60 & 70.87 &RP-2 \\\hline\hline

2.1 & 4 Robots with 12 locations & 4 & 16 & 11.86 & RP-1 \\ \hline
2.2 & 6 Robots with 12 locations & 6 & 24 & 71.50 & RP-1 \\ \hline  \hline
3.1 & Dining Philosopher 10 (no communication) & 20 &  30 &  0.25 & imp \\ \hline
3.2 & Dining Philosopher 10 (clockwise next) & 20 &  30 &  0.27 & imp   \\ \hline
3.3 & Dining Philosopher 10 (counter-clockwise next) & 20 &  30 &  0.18 & x (nor: 0.16)  \\ \hline
3.4 & Dining Philosopher 20 (counter-clockwise next) & 40 &  60 &  0.85 & x,g (nor: 0.55)   \\ \hline
3.5 & Dining Philosopher 30 (counter-clockwise next) & 60 &  90 &  4.81 & x,g (nor: 2.75)  \\ \hline\hline
4 & DPU module (local communication) & 4 &  27 &   0.42 & x   \\ \hline\hline
5 & Antenna module (local communication) & 20 & 64 & 17.21 & RP-1 \\ \hline
\end{tabular}
\begin{tablenotes}
\item[x] Satisfiable by direct fixing (without assigning any priorities)
\item[A] Nested-risk-attractor over-approximation
\item[R] State-based priority refinement
\item[RP-1] Using UNSAT core: start with smallest amount of newly introduced priorities
\item[RP-2] Using UNSAT core: start with a subset of local non-conflicting priorities extracted from the UNSAT core
\item[fail] Fail to synthesize priorities (time out $> 150$ seconds using RP-1)
\item[imp] Impossible to synthesize priorities from diagnosis at base-level (using Proposition~4)
\item[g] Initial variable ordering provided (the ordering is based on breaking the circular order to linear order)
\item[nor] Priority synthesis without considering architectural constraints (engine in~\cite{cheng:algo.priority.syn:2011})
\end{tablenotes}
\end{threeparttable}
\end{scriptsize}
\end{table}

\section{Related Work\label{sec.dps.related.work}}

Distributed controller synthesis is undecidable~\cite{PnueliFOCS90} even for reachability or simple safety conditions~\cite{Janin07On}\@.
A number of decidable subproblems have been proposed either by restricting the  communication structures between components, such as pipelined, or by restricting the set of properties under consideration~\cite{madhusudan2002decidable,madhusudan2001distributed,mohalik:2003:distributed,finkbeiner2005uniform};
these restrictions usually limit  applicability to a wide range of problems\@.
Schewe and Finkbiner's~\cite{ScheweF07a} bounded synthesis work on LTL specifications: when using automata-based methods, it requires that each process shall obtain the same information from the environment. The method is extended to encode locality constraints to work on arbitrary structures.
Distributed priority synthesis, on one hand, its starting problem is a given distributed system, together with an additional safety requirement to ensure.
On the other hand, it is also flexible enough to specify different communication architectures between the controllers such as master-slave in the multiprocessor scheduling example.
To perform distributed execution, we have also explicitly indicate how such a strategy can be executed on concrete platforms.

Starting with an arbitrary controller Katz, Peled and Schewe~\cite{DBLP:conf/cav/KatzPS11,DBLP:conf/atva/KatzPS11} propose a knowledge-based approach for obtaining a decentralized controller by reducing  the number of required communication  between components. This approach assumes a fully connected communication structure, and the approach
fails if the starting controller is inherently non-deployable.

Bonakdarpour, Kulkarni and Lin~\cite{bonakdarpour2011automated}  propose methods for adding for fault-recoveries for BIP components.
The algorithms in~\cite{bonakdarpour2008sycraft,bonakdarpour2011automated} are orthogonal in that they add additional behavior, for example new transitions, for individual components instead of determinizing possible interactions among components as in distributed priority synthesis\@.
However, distributed synthesis as described by Bonakdarpour et al.~\cite{bonakdarpour2008sycraft} on distributed synthesis is restricted to  local processes without joint interactions between components\@.

Lately, the problem of deploying priorities on a given architecture has gained increased recognition~\cite{Bonakdarpour2011distribute,bensalem2010methods}; the advantage of
priority synthesis is that the set of synthesized priorities is always known to be deployable.
\comment{
As a continuing toolchain, there are many existing works which try to deploy priorities on a given architecture~\cite{Bonakdarpour2011distribute,bensalem2010methods}, but these methods suffer from the fact that the set of priorities may be inherently undeployable. From these failures, our observation is that as priorities are control artifacts used to achieve certain specification, there can be different sets of priorities which achieve the same goal. What is required is to \emph{directly synthesize} such a priority set respecting the architectural constraints.
Therefore, our method reduces efforts in designing algorithms for priority deployment.
}

\section{Conclusion\label{sec.dps.conclusion}}

We have presented a solution to the distributed priority synthesis problem for synthesizing deployable local controllers by extending our previous algorithm for
synthesizing stateless winning strategies in safety games~\cite{cheng:vissbip:2011,cheng:algo.priority.syn:2011}\@.
We  investigated several algorithmic optimizations and validated the algorithm on a wide range of synthesis problems from multiprocessor scheduling to modular robotics.
Although these initial experimental results are indeed encouraging, they also suggest a number of further refinements and extensions.

The model of interacting components can be extended to include a rich set of data types by either using Boolean abstraction in a preprocessing phase or by using
satisfiability modulo theory (SMT) solvers instead of a propositional satisfiability engine; in this way, one might also synthesize distributed controllers for real-time systems.
Another extension is to to explicitly add the faulty or adaptive behavior by means of demonic non-determinism.

Distributed priority synthesis might not always return the most useful controller. For example, for the Dala robot, the synthesized controllers effectively shut down the antenna to obtain a deadlock-free system. Therefore, for many real-life applications we are interested in obtaining optimal, for example wrt. energy consumption, or Pareto-optimal controls.

Finally,  the priority synthesis problem as presented here needs to be extended  to achieve goal-oriented orchestration of interacting components. Given a set of goals in a rich temporal logic and a set of interacting components, the orchestration problem is to synthesize a controller such that the resulting assembly of interacting components exhibits goal-directed behavior. One possible way forward is to construct bounded reachability games from safety games.
\comment{
\footnote{Intuitively, this can achieved by a game transformation: (1) For the new arena of the transformed safety game, it is based on using a finite counter on the original reachability arena to count the number of steps which a run progresses from the initial state. (2) Set risk states to be states whose counter value is greater than the given threshold (for $k$-bounded reachability) while not in the goal set, and (3) add a self loop on each goal state (with counter less or equal to the threshold $k$) and remove existing outgoing transitions.}, our methodology is also applicable to goal-oriented orchestrations.
}

Our vision for the future of programming is that, instead of painstakingly engineering sequences of program instructions as in the prevailing Turing tarpit, designers rigorously state their intentions and goals, and the orchestration techniques based on distributed priority synthesis construct corresponding goal-oriented assemblies of interacting components~\cite{wegner1997interaction}.

\comment{
In recent years, it can be observed from the domain of industrial automation, embedded and cyber-physical systems that there is a stronger need to automatically orchestrate components to achieve or maintain certain requirements. For example, multiple hardware modules in a car shall coordinate and consume the power in an efficient or safe manner. The key ingredient is to remove the barrier of having a centralized control unit, i.e., to have decentralized control such that machines can communicate with other machines smartly without any intervention from operators. This requirement motivates our work of synthesizing distributed controllers, and it can be related to the field of distributed synthesis dated from the pioneering work of Pnueli and Ronser~\cite{PnueliFOCS90}.
}

\comment{
In this paper, we have presented methods to synthesize priorities which respect the communication architecture. The main contribution of this work has been to focus on how such distributed synthesis algorithm can be implemented symbolically, together with an implementation and evaluation over industrial application models.
In both algorithms and implementations, there are a number of directions for further work, such as scalable quantitative synthesis and the introduction of time. Our ultimate goal is to demonstrate that distributed synthesis, despite of its undecidability nature, is still practically useful in various applications.
}

\section*{Acknowledgement}

We thank Dr. Daniel Le~Berre for his kind support in guiding the use of SAT4J, and Mr. Hardik Shah for proposing the scenario listed in Section~\ref{sec.dps.introduction}.

\bibliographystyle{abbrv}
\bibliography{refs}
\newpage
\appendix

\textbf{A. Proofs for Theorem 1}


\begin{proof}
We use variable $\underline{\sigma \prec \tau}$ such that $\underline{\sigma \prec \tau}=\true$ means that priority $\sigma \prec \tau$ is included in $\mathcal{P}\cup\mathcal{P}_{d+}$. We have $|\Sigma|^2$ of such variables, and denote the set of all variables be $V_{\Sigma}$.
\begin{list1}
\item \textbf{(NP-hardness)} We have previously proven (in~\cite{cheng:hardness:2011}) that in a single player game, where \textsf{Env} is restricted to deterministic updates, finding a solution to the priority synthesis problem is NP-complete in the size $|Q|+|\delta|+|\Sigma|$ (done by a reduction from 3SAT to priority synthesis). For the hardness of distributed priority synthesis, the reduction seems to be an immediate result, as priority synthesis can be viewed as a case of distributed priority synthesis under a fully connected communication architecture. Nevertheless, as $Com$ appears in distributed priority synthesis and does not appear in normal priority synthesis, we also need to consider time used to construct the fully connected architecture, which is of size  $|C|^2$. Notice that $|C|$ is not a parameter which appears in the earlier result. This is the reason why we need special care to constrain $|C|^2$ to be bounded by $|Q|+|\delta|+|\Sigma|$. With such constraint, as (1) the construction of fully connected architecture is in time polynomially bounded by $|Q|+|\delta|+|\Sigma|$, and (2) the system is the same, we obtain a polynomial time reduction.

    \emph{[Formal reduction]} For the reduction from priority synthesis (environment deterministic case) to distributed priority synthesis, given $\mathcal{S}$, we construct the fully connected architecture $Com$. As $|Com| = |C|^2$, based on the assumption where $|C|^2 < |Q|+|\delta|+|\Sigma|$, the time required for the construction is polynomially bounded by $|Q|+|\delta|+|\Sigma|$.
    \begin{list1}
        \item \textbf{($\Rightarrow$)} Assume $\mathcal{P}_{+}$ is the set of priorities from priority synthesis such that $(C , \Sigma, \mathcal{P}\cup\mathcal{P}_{+})$ is safe. Then for the translated problem (distributed priority synthesis with fully connected architecture), all priorities in $\mathcal{P}_{+}$ are deployable, so $\mathcal{P}_{+}$ is also a solution for the translated problem.
         \item \textbf{($\Leftarrow$)} The converse is also true. 
    \end{list1}
\item \textbf{(NP)} Nondeterministically select a subset of $V_{\Sigma}$ and assign them to \true (for others set to \false), and such a subset defines a set of priorities. We need to check the corresponding priorities satisfies three conditions of distributed priority synthesis (Definition~5).
    \begin{list1}
\item The first condition can be checked by computing the transitive closure and is in time cubic to $|\Sigma|^2$.
\item The second condition can be checked by using a forward reachability analysis (from initial configuration) to compute the set of reachable states, and during computation, check if any bad state is reached. During the reachability analysis, every time we try to add a $\sigma$-successor $c'$ from a configuration $c$, we check if there exists a priority $\sigma \prec \sigma'$ where $\underline{\sigma \prec \tau}$ is evaluated to $\true$ and $\tau$ is also enabled, such that $\tau$ blocks the adding of $c'$ to reachable set. The overall time for the analysis is linear to $|Q||\delta||\Sigma|^2$.
\item For the last condition, we check if  $\underline{\sigma \prec \tau} = \true$, for all $C_i$ where $\tau\in\Sigma_i$ and $C_j$ where $\sigma\in\Sigma_j$, $C_j \informs C_i \in Com$.
    \begin{list1}
        \item Each checking involves at most $|C|\times|C|$ pairs. There are at most $|\Sigma|^2$ variables that need to be checked.
        \item Each pair is checked in time linear to $|Com|$, where $|Com|$ is bounded by $|C|^2$.
        \item Therefore, the total required time for checking is bounded by $\mathcal{O}(|C|^4|\Sigma|^2)$.
        \item As $|C|^2 < |Q|+|\delta|+|\Sigma|$, the total required time for checking is polynomially bounded by  $|Q|+|\delta|+|\Sigma|$.
    \end{list1}
\item In addition, we also check if the selected set contains $\mathcal{P}$, which is done in time polynomially bounded by $|\Sigma|^2$.
    \end{list1}


\end{list1}
\end{proof}

\textbf{B. Soundness of the SAT Resolution in the Fixing Process}

Concerning correctness of the whole fixing algorithm, the key issue is whether it is possible for the SAT resolution to create a set of priorities which is unable to block the entry to the nested-risk-attractor (if it is unable to do so, then the algorithm is incorrect). Although our algorithm is performed symbolically, it is appropriate to consider each location separately (as if there is no symbolic execution).

For a control location $s$ where $s$ is within the source of $\mathcal{T}_f$ returned from Algorithm~\ref{algo.prioritysyn.nested.attractor} (recall in Section~\ref{subsec.dps.core.algorithm.game} we call $s$ an error point), we denote the set of its outgoing transitions as $T_s$. Recall that for each transition in $T_{s}$, it represents a unique selection (execution) of an interaction. We use $\Sigma_s \subseteq \Sigma$ to represent the set of corresponding interactions in $T_s$. $\Sigma_{s}$ can be partitioned to $\Sigma_{s,bad}$ and $\Sigma_{s,good}$, where  $\Sigma_{s,bad}$ are interactions which enter the nested-risk-attractor, and $\Sigma_{s,good}$ are interactions which keep out from the nested-risk-attractor. Notice that the size of $\Sigma_{s,good}$ is at least~$1$ (otherwise, $s$ shall be added to the nested-risk-attractor by the inner while-loop of Algorithm~\ref{algo.prioritysyn.nested.attractor}).

We now prove that: If the SAT solver returns a solution (it is also possible to return unsatisfiable, but then we just report no fix-solution is generated and continue the \textsf{DPS} algorithm), then for all error point $s$, each $\sigma \in \Sigma_{s,bad}$, there exists $\tau \in \Sigma_{s,good}$ such that $\sigma \prec \tau$ is in the synthesized priority set (Then at $s$, as $\tau$ is enabled, $\sigma$ is guaranteed to be blocked).

\begin{proof}
The proof proceeds as follows.
\begin{enumerate}
    \item (Guaranteed by Algorithm~\ref{algo.prioritysyn.nested.attractor}, line 9) As $s$ is not inside the nested-risk-attractor, $\forall \sigma \in \Sigma_{s,bad}, \exists \Sigma_{\sigma} \subseteq \Sigma_{s}\setminus \{\sigma\}$ such that $\forall \tau \in  \Sigma_{\sigma}, \Vis^{\tau}_{\sigma} = \true$. Therefore, each bad interaction will have at least one fix candidate.
    \item (Definition of staying outside nested-risk-attractor) $|\Sigma_{s,good}| \geq 1$. Therefore, at least one edge is a true escape, whose destination is outside the nested-risk-attractor.
    \item (Assume contradiction) Assume that when SAT solver claims satisfiable, but from the return information, exists $\sigma \in \Sigma_{s,bad}$ where no priority $\sigma \prec \tau$, where $\tau \in \Sigma_{s,good}$.
    \item (Consequence) From~1 and~3, then exists $\sigma_{bad1}\in \Sigma_{s,bad}$, where SAT solver returns priority $\sigma \prec \sigma_{bad1}$.
    \item (Violation: Case 1) From~1, then $\sigma_{bad1}$ also has a fix candidate. If the SAT solver returns $\sigma_{bad1} \prec \sigma_{good}$, where $\sigma_{good} \in \Sigma_{s,good}$, then due to transitivity (SAT clause Type 4), then $\sigma \prec \sigma_{good}$ shall be returned by the SAT solver. Contradiction.
    \item (Violation: Case 2) Otherwise, SAT solver only returns $\sigma_{bad1} \prec \sigma_{bad2}$, where $\sigma_{bad2} \in \Sigma_{s,bad}$. From this, the chain $\sigma \prec \sigma_{bad1} \prec \sigma_{bad2} \ldots $ which consists only $\Sigma_{bad}$ continues. However, this priority chain will either stop by having an element in $\Sigma_{good}$ (then it jumps to Case 1 violation), or it move to cases where a repeated element (which occurred previously in the chain) eventually reappears.  Notice that if the chain does not jump an interaction $\sigma'\in \Sigma_{s,good}$, eventually it has to use a bad interaction repeatedly, as the chain $\sigma \prec \sigma_{bad1} \prec \sigma_{bad2} \ldots $ can have at most $|\Sigma_{s,bad}|$ ``$\prec$" symbols (because every element in $\Sigma_{s,bad}$ needs to be fixed, based on~1), but for that case, there are $|\Sigma_{s,bad}| + 1$ elements in the chain, so Pigeonhole's principle ensures the repeating of a bad interaction $\sigma_{bad.r}$. When it reappears, then there is an immediate violation over SAT clause Type 3 (irreflexive), as transitivity brings the form $\sigma_{bad.r} \prec \sigma_{bad.r}$, which is impossible. 
    \item Therefore, the assumption does not hold, which finishes the correctness proof.
\end{enumerate}
\end{proof}

\end{document}